# High-temperature superconductors of the family (RE)Ba$_2$Cu$_3$O$_{7-\delta}$ and their application (Review Article)

S. I. Bondarenko,[a] V. P. Koverya, A. V. Krevsun, and S. I. Link

*B. Verkin Institute for Low Temperature Physics and Engineering, NAS Ukraine, 47 Nauka Ave., Kharkov 61103, Ukraine*

This review article is a commemoration of the 30th anniversary of the discovery of YBa$_2$Cu$_3$O$_{7-\delta}$ high-temperature superconductors (HTSCs). As a result of this discovery a family of (RE)Ba$_2$Cu$_3$O$_{7-\delta}$ (RE stands for "rare earth") HTSCs has found great practical use. The review article consists of a brief history of how YBa$_2$Cu$_3$O$_{7-\delta}$ was conceived and five sections describing the family of compounds: crystallography, phase diagrams, manufacturing techniques, main superconducting properties, and fields of application.

## 1. Introduction. A brief history of how the YBa$_2$Cu$_3$O$_{7-\delta}$ compound was created

In 1986 Swiss physicists Bednorz and Müller published an article[1] detailing their experimental discovery of a new type of superconductor based on (La, Ba)$_2$CuO$_4$ with a transition temperature $T_c$ to the superconducting state of about 30 K. This temperature exceeded the transition temperature of the compound Nb–Ge ($T_c \approx 23$ K), which held the record at the time, by only 30%. It was yet another discovery that turned out to be astonishing. This new superconductor was prepared using three non-superconducting components: lanthanum, barium and copper oxides, and was a typical ceramic. This was unusual and highly interesting from the perspective of superconductivity physics, but not in terms of its possibilities for application. In order for a real revolution to occur in the eyes of superconductivity researchers there had to be a superconductor with a significantly higher transition temperature. Such a revolution was accomplished by a group of researchers from the University of Houston in the U.S., led by Paul Chu. In January 1987 he patented a new superconducting compound YBa$_2$Cu$_3$O$_{7-\delta}$, a ceramic with $T_c \sim 90$ K (the value of $\delta$ is discussed below). This moved superconductivity out of the category of expensive and often unique "helium" technologies that existed from the moment superconductivity was discovered in 1911 and functioned at a temperature of about 4 K, into a much more accessible category of "nitrogen" technology implemented using a cheap liquid nitrogen as a coolant with a boiling temperature of about 77 K. In this same year an article by Paul Chu[2] containing the results of the first experimental studies on the properties of this revolutionary compound (Fig. 1) was published.

In the article, the authors report on another record parameter of this compound, which is the value of the second critical magnetic field $H_{c2}$ at $T \rightarrow 0$ K. According to their estimates $H_{c2}(0) \approx 180$–$200$ T. If these values were to be experimentally confirmed, such a superconductor would allow for the creation of sources of constant magnetic fields with unprecedented magnitude. In turn, this promised a technological revolution, particularly in the field of creating small powerful electric cars and power lines without resistive energy losses.

The newly discovered opportunities for using the high-temperature superconductor (HTSC) YBa$_2$Cu$_3$O$_{7-\delta}$ gave rise to the sort of enthusiasm in the scientific and technical communities of various countries that probably has not been seen since the developments in atomic energy and semiconductor transistors. First and foremost, this touched those countries in which a lot of attention had traditionally been devoted to superconductivity: the United States, the Soviet Union, some countries in Europe, and Japan. In this regard it is enough to recall the mass Moscow physics seminar that took place in the summer of 1987 at the Institute for Physical Problems of the USSR Academy of Sciences, and was dedicated to the discovery of high temperature superconductivity. This seminar was unparalleled in the scientific history of the country. At the event, Moshchalkov demonstrated the magnetic levitation of a HTSC sample in liquid nitrogen vapor, directly on the stage of the IPP assembly hall. Similar events took place in other countries, such as the assembly of 2000 researchers in New York on March 18th, 1987. New scientific journals on superconductivity were published in the USSR and abroad. The governments of a number of countries established programs dedicated to the study of HTSCs. The USSR program was carried out under the supervision of the State Committee for Science and

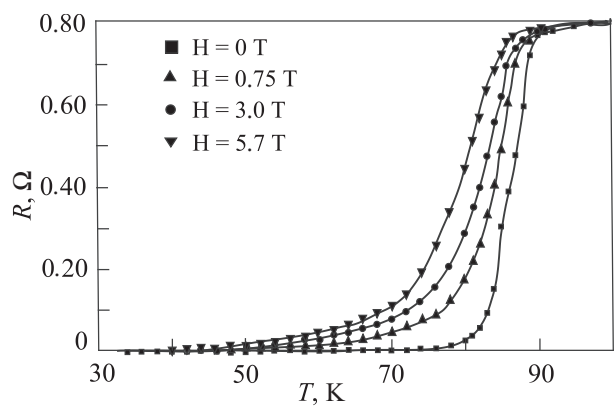

Fig. 1. The first dependences of the electrical resistance $R$ of a YBa$_2$Cu$_3$O$_{7-\delta}$ ceramic sample on the temperature at different values of the external magnetic field with $H = 0$–$5.7$ T,[2] published in 1987.

Technology starting from 1987 until the dissolution of the Soviet Union. The program envisaged the allocation of significant financing to the study of HTSCs on a competitive basis. In particular, government funds were used to purchase modern technological and computer equipment abroad and build institutions and labs. For example, funds in the amount of half a million U.S. dollars were competitively allocated for research on weak superconductivity alone—a field where ILTPE was a leading institution. The projects provided for the theoretical and experimental study of HTSCs throughout the country for those applied uses that had traditionally been implemented with the help of helium superconductors: weak superconductivity and superconducting electronics, high-current superconductivity, devices for creating high magnetic fields, cryogenic equipment for creating the necessary low-temperature level, and materials science studies for creating high-quality superconductivity materials. The scientific cooperation between the researchers of the ILTPE and the Institute for Single Crystals (ISC), organized by the directors of those institutions, B.I. Verkin and V.P. Seminozhenko, was of a decisive importance to HTSC research. In the first years of studies pertaining to HTSCs the majority of the $YBa_2Cu_3O_{7-\delta}$ ceramic samples came to ILTPE from ISC technologists. During their five years of cooperation a number of fundamental results were obtained (see below). As such, one of the most important fundamental problems for that period of research was the question of whether or not electron pairing, long-range order, and a bandgap existed in $YBa_2Cu_3O_{7-\delta}$ ceramic samples. Positive answers to these questions were given for the first time ever in 1987 by independent researchers at ILTPE in the USSR[3–6] and the USA.[7–9] The appearance of a step along the current-voltage (I-V) curve of a $YBa_2Cu_3O_{7-\delta}$–$YBa_2Cu_3O_{7-\delta}$ point junction when it was exposed to microwave radiation at a frequency of $\nu = 10\,GHz$ (Ref. 3) served as evidence of pairing. The voltage $V$, at which this step occurred along the I-V curve, corresponded to the well-known Josephson relation

$$2eV = h\nu, \quad (1)$$

wherein $2e$ and $h$ are the charge of the Cooper pair and Planck's constant, and $\nu$ is the frequency of the electromagnetic radiation. The long-range order was established by observing the quantum interference in two different types of HTSCs by a quantum interferometer at variable[4] and constant[5] currents. The feasibility of Josephson effects under weak coupling such as a HTSC clamping point contact[3,8] was experimentally proven at the same time, and the magnitude of the bandgap in the $YBa_2Cu_3O_{7-\delta}$ ceramic was measured.[6]

A period of intense study of cuprate superconductors began all over the world. It soon became clear that we are dealing not only with the discovery of a superconductor that has a high critical temperature, but also with a whole other type of superconductivity, which was later referred to as "unusual." This unusuality will be discussed in the next sections of this review.

The high activity of research on HTSCs was accompanied and continues to be accompanied by the ongoing publication of an enormous amount of scientific literature. There are indications that the number of such publications considerably exceeds the number of articles about superconductivity that were released in the first 75 years that superconductors existed before the discovery of HTSCs. In this review we limit ourselves to studying articles related mainly to the first HTSC $YBa_2Cu_3O_{7-\delta}$ and the associated compounds from the $(RE)Ba_2Cu_3O_{7-\delta}$ family with a critical temperature above 77 K. Even in this case, due to the massive array of information available on these compounds we are forced to limit ourselves only to the most important (from our point of view) information (mainly in the form of experimental results) on the superconducting properties of $(RE)Ba_2Cu_3O_{7-\delta}$ that is most closely associated with their application to science and technology.

## 2. The crystallography of $(RE)Ba_2Cu_3O_{7-\delta}$ (RE: rare earth Y, Nd, Sm, Eu, Gd, Dy, Ho)

The first feature of the considered superconducting compounds, which separates them from earlier-known low-temperature superconductors (LTSC), is their significantly more complex chemical elemental composition that does not contain any of the previously known LTSC superconducting elements. The HTSCs of this family are complex layered metal oxide compounds. The unit cell of the $YBa_2Cu_3O_{7-\delta}$ compound crystal lattice[10,11] is similar in structure to three unified perovskite unit cells ($K_2NiF_4$) and is composed of yttrium (Y), barium (Ba), copper (Cu), and oxygen (O) atoms (Fig. 2).

The unit cell of superconducting compounds with other lanthanides (Nd, Gd, Sm, Dy, Ho, Eu, Er, Tm, Yb, Lu) is similar to the $YBa_2Cu_3O_{7-\delta}$ cell when yttrium is replaced by the corresponding lanthanide. The atoms of these elements, except for yttrium, have a magnetic moment in the range of (3.62–10.5) $\mu_B$. The critical temperatures of these compounds are close to each other and differ by no more than 3 K.[14] The dimensions of the unit cells for the three most common superconducting compounds are given in Table 1.

Lanthanides are metals. The majority of them have hexagonal cells; europium has a body-centered cubic lattice, whereas ytterbium is cubic. Two different modifications of neodymium are known to us.[15] The cell of the initial (before oxygen doping) compound based on yttrium ($YBa_2Cu_3O_6$) with oxygen vacancies has a tetragonal crystal lattice. During the thermal synthesis of the compound in the presence of oxygen, vacancies are filled by a certain number of oxygen atoms (see the section on p. 3) and its lattice becomes orthorhombic, thus leading to the appearance of electrical conductivity in the crystal and its transition to the superconducting state with $T_{c\,max} \approx 93$ K given optimal oxygen doping ($\delta = 0.06$).[15]

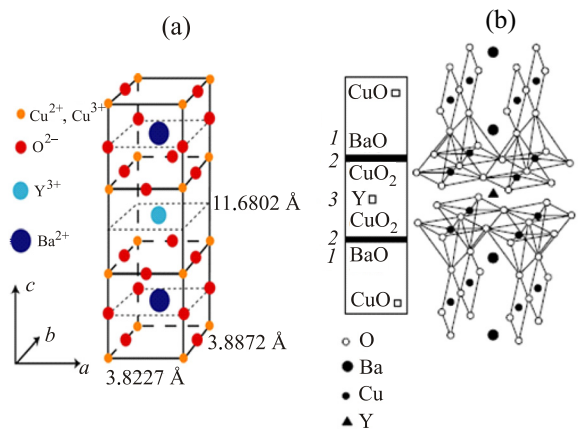

Fig. 2. Images of the $YBa_2Cu_3O_{7-\delta}$ compound unit cell: (a) includes the unit cell dimensions along its main axes $a,b,c$;[12] (b) is the octahedral view of the cell including a conventional image to the left of it showing the position of the dielectric (1), oxygen-deficient (3) and superconducting planes $CuO_2$ (2).[13]

TABLE 1. The dimensions of the three most common superconducting compounds.

| Compound | $a$, Å | $b$, Å | $c$, Å |
|---|---|---|---|
| YBCO[13] | 3.8227 | 3.8872 | 11.6802 |
| GdBCO[13] | 3.837 | 3.677 | 11.786 |
| NdBCO[14] | 3.911 | 3.913 | 11.725 |

The properties of the orthorhombic phase are influenced by twins having a distance of $\sim 1000$ Å (Ref. 16) between their walls. The twin walls play the role of pinning centers: the hysteresis loop of magnetization for crystals with twins is much larger than that of crystals without twins.[17] In single crystals wherein twin walls are oriented in one direction, $T_c$ is higher by 1...2 K in comparison to $T_c$ samples with different directions of these walls.[18] This, in particular, could explain the increased value $T_c \sim 93$ K in single crystals in comparison to $T_c = 91.92$ K in polycrystals. An example of the twin wall structure is shown in Ref. 19. In this example various coordination relationships between copper and oxygen ions are considered: a $CuO_2$ dumbbell, a $CuO_4$ square, $CuO_5$ tetrahedron, $CuO_6$ octahedron.

A typical morphological feature of such crystals is their layering. The layer is a single crystal with a thickness equal to size $c$ of the unit cell. The distance between layers, comparable with the dimensions of the compound unit cell, allows for the formation of the interlayer Josephson junction. There is an opportunity for natural Josephson junctions to exist, and for the internal Josephson effect (IJE) to be observed in a crystal containing many successively connected Josephson junctions.[20]

As shown by the study of the compound $YBa_2Cu_3O_{7-\delta}$ (Ref. 21) the unit cell of a superconducting crystal has a remarkable feature. The electrical conductivity and superconductivity of the crystal are largely determined by the high electroconductive properties of planes (2) (Fig. 2), the two planes closest to the center of the cell. The high electrical conductivity of the structure, similar to that of plane (2), was confirmed by the experiment on the interface between the copper film and the CuO single crystal.[22,23]

Another important parameter that determines these superconducting properties is the Cu–O bond length in the (2) plane, which must lie in the interval between 0.190 and 0.197 nm. Copper atoms can also be bound with oxygen atoms located in adjacent layers, however these bonds should be much longer and exceed 0.22 nm. In other words, superconducting cuprates have unequal Cu–O bonds: strong bonds in the plane of each $CuO_2$-layer and much weaker bonds in a direction perpendicular to these layers. A more detailed description of the complex electronic processes involved in the planes of the lattice with copper ions is beyond the scope of this review.

## 3. Phase diagrams of the $YBa_2Cu_3O_{7-\delta}$ compound

One of the peculiarities of the compounds up for consideration that distinguish them from LTSCs is the extremely high sensitivity of their properties to the content of oxygen.[11] This is clearly visible in the phase diagram of the $YBa_2Cu_3O_{7-\delta}$ compound (Fig. 3).

It can be seen in Fig. 3 that for $YBa_2Cu_3O_{6+x}$ (this is another designation of the same $YBa_2Cu_3O_{7-\delta}$ compound, used by the authors of Ref. 14) at $x > 0.4$ the compound is electrically conductive and superconducting, whereas at $x < 0.4$ it becomes a Mott insulator with long-range antiferromagnetic order.

Even in early studies of the Hall effect in these HTSCs it was established that they have "hole" conductivity.[24] The charge carriers in these compounds are mainly holes, not electrons. Oxygen ions play the role of the hole-dopant. Therefore these compounds are classified as so-called "hole" superconductors, as opposed to LTSCs in which electron conductivity is predominant and are referred to as electron superconductors. It was experimentally established that for superconductivity to occur in $YBa_2Cu_3O_{7-\delta}$ it is necessary for the valence of copper in $CuO_2$-layers with collectivized electrons to differ slightly form $+2$ and be in the range between 2.05 and 2.25.[25,26]

Figure 4 shows that at $0.06 < x_h < 0.275$ the compound $YBa_2Cu_3O_{7-\delta}$ is electrically conductive and is a superconductor. At the same time a compound with $x_h \approx 0.17$ has the highest critical temperature. The maximum values of $T_c$ are attained at optimum (in terms of reaching maximum $T_c$) oxygen content in the superconducting phase.[27] Since for $YBa_2Cu_3O_{7-\delta}$ $(T_c)_{max} = 93.5$ K, then $\delta_{opt} = 0.06$, and in the notation of the authors of Ref. 14, $x_{opt} = 0.94$. Thus, using a different level of oxygen doping for the compound makes it possible to vary its superconducting properties over a wide range. The phase diagrams of other $(RE)Ba_2Cu_3O_{7-\delta}$ are qualitatively similar to the diagrams of the $YBa_2Cu_3O_{7-\delta}$ being considered, and their maximum critical temperatures differ little from the values for $YBa_2Cu_3O_{7-\delta}$.

Impurities are almost always present in prepared HTSC materials. Impurities infiltrate the crystallizing system from the initial reagents containing accompanying elements and anionic groups, as well as from the containers in which the crystallization is carried out. Information on the physiochemical state diagrams of the systems with impurities, as well as on the effect the impurities have on HTSC properties is necessary for a purposeful optimization of the HTSC preparation methods with high superconducting parameters [Ref. 28]. It was established for a number of impurities that their presence in the ceramics at concentrations of $10^{-4} < n < 10^{-1}$ leads to an increase in the transition temperature $T_c$ by 1–5 K and to a decrease in the width of the superconducting transition $\Delta T_c$. Such cationic impurities include Ti, Ag, Pd, La. According to the authors of Refs. 29 and 30 the presence of carbon leads to a slight decrease in $T_c$. At the same time, according to Ref. 31, the synthesis of $YBa_2Cu_3O_{7-\delta}$ in the presence of organic compounds makes it possible to increase $T_c$ to 113–115 K.

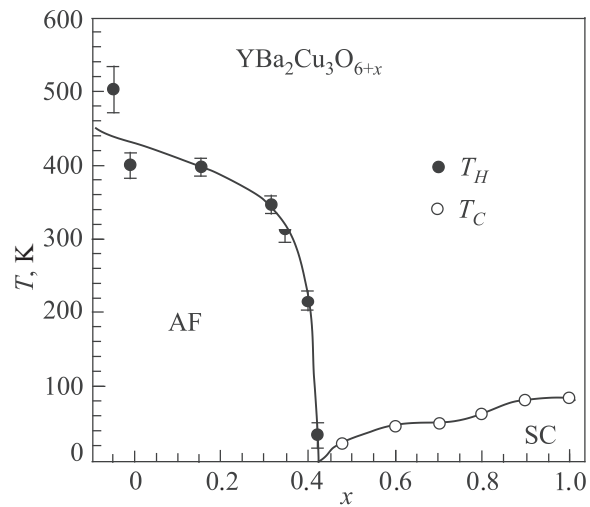

Fig. 3. The dependence of the critical temperature $T_c$ and the Néel temperature $T_N$ of the $YBa_2Cu_3O_{6+x}$ compound, on $x$.[14]

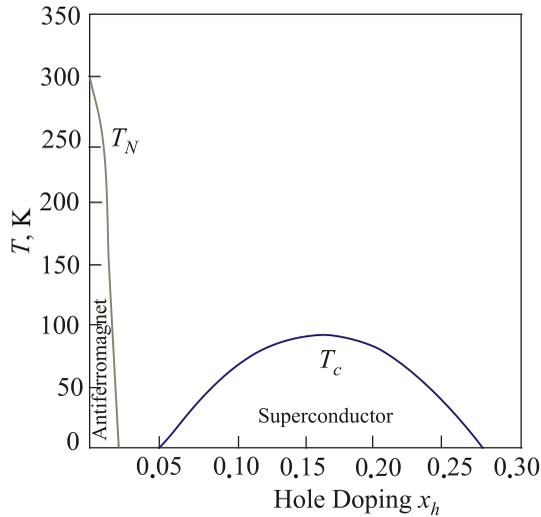

Fig. 4. A fragment of the $YBa_2Cu_3O_{7-\delta}$ phase diagram, showing the regions of existence for the two main phases of the compound (superconducting and antiferromagnetic), as well as their transition temperatures $T_N$ and $T_c$, as functions of the number of holes $x_h$ per Cu ion.[26]

## 4. The technology involved in preparing (RE)Ba₂Cu₃O₇₋δ samples

This review considers the technology involved in preparing the three main types of HTSCs: thin films, extended superconductors (wires), and bulk (volumetric) formations in the form of plates, cylinders, and discs. This set of HTSC types is determined by the main fields of superconductor application and does not differ from the types of LTSCs. At the same time the technology of their preparation is fundamentally different from the technology involved in manufacturing LTSCs.

### 4.1. Films and extended superconductors

At present, the following areas are identified as priority tasks when it comes to the technology involved in obtaining (RE)Ba₂Cu₃O₇₋δ (REBCO) films and studying their properties:[32–56]

- developing effective methods for obtaining highly-oriented HTSCs and intermediate (buffer) non-superconducting films between the substrate and the HTSC film, as well as the fundamental study of epitaxial growth mechanisms and grain boundary structures of the films prepared using both physical and chemical methods;
- producing superconducting films with high values of critical currents ($I_c$) on flexible metal strips of nickel and alloys thereof, coated with a buffer film and textured by rolling and annealing;
- testing the technology for obtaining HTSC films with a large number of pinning centers to improve their performance in high magnetic fields;
- studying the properties of HTSC films at different orientations relative to the external magnetic field;
- investigating the impact mechanical stresses (temperature, strain) have on the morphology and electrical properties of superconducting films and coatings.

To date, vast experience related to the methods of obtaining primarily $YBa_2Cu_3O_{7-\delta}$ films, which can be divided into physical and chemical, has been accumulated. The physical methods involved can be divided by the type of action the target material is subjected to, which is either ionic sputtering (ion-plasma and ion-beam deposition) or thermal evaporation (pulsed laser and electron-beam evaporation, laser molecular beam epitaxy).[44,57] The physical methods make it possible to obtain thin films of high quality with top-notch physical characteristics, as well as to perform a layer-by-layer synthesis of new structures (structural design), thus literally "collecting" the film at the level of atomic planes. However, these physical methods are practically inapplicable to the process of manufacturing extended samples (wires).

The chemical methods for producing HTSC films are based on precursors (starting materials) that are deposited on substrates, the thermal treatment of which as a result of chemical reactions leads to the formation of a film that has the required composition.[58] The method of application and the aggregate state of the precursors are used to distinguish between chemical solution deposition (CSD), and chemical vapor deposition (CVD).[45] The wide use of chemical methods is due to a number of advantages they have over physical techniques, namely: composition universality; the ability to apply films that are uniform in composition and thickness to large surfaces and extended conductor ribbons; high performance, simplicity, and inexpensive equipment.[59] Among the most widely used chemical procedures for producing HTSC films are methods such as chemical solution deposition,[60] sol-gel,[61] as well as metal-organic chemical vapor deposition (MOCVD)[62] are the most noteworthy.

Quality superconducting REBCO films (see also Sec. 5.3) are obtained under the condition that their elemental composition be strictly observed, and that the necessary crystal structure with optimum oxygen content and minimal disorientation of its grains is present. The superconducting properties of the films are significantly influenced by the substrates used for their deposition.[44]

### 4.1.1. HTSC substrates

The substrates of deposited films have the following requirements:

- the crystal lattices of the substrate and the film must correspond;

- the thermal expansion coefficients of the film and substrate must be consistent;
- chemical stability with respect to the film materials and manufacturability (cheapness, workability and operational characteristics, the necessary dimensions, twin minimum);
- additional requirements related to the features of film application (for example, suitable dielectric properties).

A high-quality $YBa_2Cu_3O_{7-x}$ film has crystalline cell parameters that are close to those indicated in Table 1, and the linear thermal expansion coefficient in the $ab$-plane (001) is $\lambda = 1.1 \times 10^{-5} \, K^{-1}$.[63]

Modern methods of obtaining HTSC films allow one to manufacture high-quality $YBa_2Cu_3O_{7-\delta}$ films with the following characteristics:[44,64–66] surface roughness 10–20 nm, resistivity $\rho(300 \, K) \approx 350$–$400 \, \mu\Omega$ cm, resistance ratio $\rho(300 \, K)/\rho(100 \, K) \approx 3$–3.3; microwave surface resistance $R_s$ (77 K, 10 GHz) $\approx 400$–500 $\mu\Omega$; a critical temperature of 87–90 K with a temperature transition width that is a fraction of a degree; the critical current density $j_c$ is about $10^6 \, A \, cm^{-2}$ at $T = 77 \, K$. The substrates that are the most suitable in terms of structure but not in terms of dielectric properties are: $SrTiO_3$ (STO) with a cubic lattice ($a = 3.905$ Å),[44,67] $LaAlO_3$ with a distorted cubic lattice ($a = 3.792$ Å),[68] MgO with a cubic lattice ($a = 4.203$ Å), and $ZrO_2$ (YSZ) yttria-stabilized zirconia with a cubic lattice ($a = 5.16$ Å). Sometimes substrates from other materials are used (for example, sapphire with buffer layers of $CeO_2$, Ag, MgO[69,70]) on which epitaxial buffer layers are preliminarily grown to eliminate significant lattice discrepancies and/or chemical interactions between the film and the substrate. In some cases buffer layers are used even on "structurally suitable" substrates, for example when a $CeO_2$ buffer layer is used on a $LaAlO_3$ substrate to reduce the probability of film growth in the direction of the axis $a$. The use of substrates prepared from tetragonal solid solutions such as $Pr_{1+x}Ba_{2-x}Cu_3O_z$, in which there is no twinning,[71] as well as dielectric single crystals $Nd_{1.85}Ba_{1.15}Cu_3O_z$, characterized by a high degree of rhombicity and the absence of a tetra-ortho transition, is promising. They have close thermal expansion coefficients (TEC), their parameters are in good agreement with HTSC films, and they have a low degree of oxygen nonstoichiometry.[72]

When HTSC films are used in microwave technology their dielectric properties such as the dielectric constant $\varepsilon$, and the dielectric loss tangent tg $p$, are very important substrate parameters in the required range of operating frequencies. In such a case, in addition to the requirement that the dielectric properties be reproducible for the entire batch of utilized substrates and that they be homogeneous with respect to area, the following limitations also apply: tg $p < 10^{-4}$ and $\varepsilon \leq 10$ at frequencies greater than 10 GHz. Reference 7 contains a detailed analysis of the dielectric properties of various substrates for use in microwave electronics. The properties of the substrate materials suitable for depositing REBCO films are also provided in many other articles, including Refs. 44, 57, and 62.

Recently, due to the need to manufacture long (hundreds of meters) HTSC conductors, textured tapes based on Ni and N-W with buffered, protective, and HTSC films chemically deposited on them have been used.[59,61]

Let us briefly consider the basic technologies for obtaining HTSC films, the problems that arise during the manufacturing process, solutions to these problems, as well as the merits and shortcomings of each method.

### 4.1.2. The physical methods of manufacturing films

Figure 5 (Ref. 57) shows the phase stability diagram of the $YBa_2Cu_3O_{7-\delta}$ film and the typical modes of physical processes involved in HTSC film preparation. The orthorhombic phase of $YBa_2Cu_3O_{7-\delta}$ is stable in the right part of the darkened region on Fig. 5, and the dashed line separates the tetragonal and superconducting orthorhombic phases. In this case the preparation of the film occurs according to ion-plasma sputtering. The plasma consists of argon and oxygen ions. In a vacuum chamber they are referred to as working gases. The deposited film has a tetragonal lattice. In order to obtain the orthorhombic phase the film is processed post-growth without opening the chamber (in situ). Paths $A$ or $D$ (Fig. 5) correspond to the in situ post-processing of the deposited film. In situ sputtering technology usually consists of two stages: the deposition of the film at a high substrate temperature (650–800 °C), depending on the working pressure, and post-growth exposure in an oxygen atmosphere, over the course of which the film goes through a phase transition from the tetragonal to the superconducting orthorhombic phase. More information on the post-growth processing can be found in Sec. 4.1.3.

When using fast inert gas ions (usually $Ar^+$) ion-plasma and ion-beam technologies are distinguished from one another. In the first case the target with the substrate is in the discharge chamber, and the target is used as a cathode. Diode, triode and magnetron sputtering systems are differentiated from one another according to their construction, and high-frequency and DC sputtering[69,74] are differentiated based on the type of electric current supplying the discharge. In any variant the ion-plasma sputtering must take into account the various factors affecting the stoichiometry of the resulting film, namely:

- the change in the composition of the oxide target during the process of its sputtering (due to the different sputtering speed of its components and their diffusion from the bulk to the near-surface layer);
- different angular distribution of the sputtered particles;

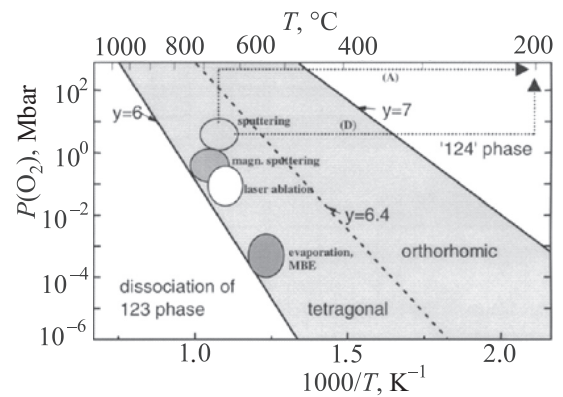

Fig. 5. A phase stability diagram for the $YBa_2Cu_3O_y$ film ($y = 7-\delta$) film in oxygen pressure-substrate temperature coordinates. The method of deposition and oxygen content is indicated.[57]

- the redistribution of the film as a result of its bombardment by high-energy oxygen ions;
- the change in the composition of the film due to the atoms' different adhesion coefficients to the substrate, which is especially pronounced at a substrate temperature greater than 400 °C.

Possible ways of solving this problem of non-stoichiometry (according to the content of metallic elements) of the film are:[44]

- depositing the films in a steady-state spraying mode;
- spraying at very high working gas pressures and a system design solution that allows for the minimization of discharge voltage;
- using non-stoichiometric targets and/or using composite or multiple targets;
- placing the substrate according to off-axis geometry with respect to the flow of charged particles.

At present, magnetron sputtering is used in most of the cases, since it allows researchers to obtain a denser plasma at lower discharge voltage and higher film deposition rates (units of mono-layers per second). The pressure of the working gas in this case is maintained in the range of 1–10 Pa. Unfortunately, the problem of negative oxygen ions can so far only be solved by reducing the film deposition rates. It should also be noted that calibration control and deposition rates are made complicated by the presence of oxygen in the vacuum chamber.

In ion-beam deposition[57,75] the target is sprayed by an ion beam extracted from the plasma source. The sputtering of the target is carried out at a lowered ($10^{-1}$–$10^{-2}$ Pa) working gas pressure, the substrate is blown over by a flow of oxygen. Since the oxygen pressure is several orders of magnitude less than in ion-plasma sputtering systems we face the problem of an oxygen deficiency in the growing film. The merit of this method includes the low content of impurities in the film, good controllability of the sputtering conditions, and the reproducibility of film properties. The accuracy of target composition inheritance depends on the temperature of the substrate, which determines the processes of condensation, oxidation, and desorption of the components belonging to the sputtered substance.

Pulse laser deposition (PLD), or ablation, is one of the most widely used methods for producing HTSC films and is based on applying thermal action to the target.[44] A laser beam with an energy of 1–3 J cm$^{-2}$ focused on one spot (energy density $>10^{7}$–$10^{8}$ W cm$^{-1}$) with a pulse repetition rate of several Hz acts on the rotating target of the HTSC material. As a result of this action there is a strong local heating at the laser beam incidence point, and the formation of a vapor cloud composed of the target material that condenses on the substrate. Pulse laser deposition has several advantages. Unlike other methods, the thermal effect on the target is easily controlled regardless of the pressure and composition of the gas mixture, simple control over stoichiometry at high energy densities is possible, as are high rates of deposition and ease of control over the process. However, the film area does not exceed 1 cm$^2$ and strong effects on the morphology of the sample are present during deposition; the sample surface acquires submicron droplets of target material. In order to increase the deposition area the sample is usually rotated, whereas in order to reduce the morphology effect on the film surface the substrates are removed and positioned according to off-axis geometry. In Ref. 76 a fast-filtration method was successfully implemented, which allowed for smooth uniform epitaxial YBa$_2$Cu$_3$O$_{7-\delta}$ films to be obtained.

In contrast to the typical processes of ionic sputtering and laser ablation, which use a single target, thermal co-precipitation or molecular-beam epitaxy use individual sources for each of the metals contained in the film.

Reference 77 provides a principle layout for a thermal co-precipitation system. The sputtering is performed in high vacuum of $\sim 10^{-5}$ mbar, which allows for the use of highly sensitive sensors to control the deposition rate and thickness, which therefore allows for highly accurate control over the stoichiometry, exceeding 1%. Another advantage of such low operating pressure is the ability to obtain a stable tetra phase at a lower substrate temperature (650 °C). In the presented layout the rotating substrate periodically leaves the pocket with a higher ($\sim 2$ orders of magnitude) oxygen pressure into the sputtering region, which allows for the alternation between the deposition and oxidation processes and produces highly thin and quality films.

### 4.1.3. Post-growth processing of films

As already noted, quality YBa$_2$Cu$_3$O$_{7-\delta}$ films have a crystal structure given an optimum oxygen content ($7-\delta = 6.94$), which is determined by the deposition process and/or the post-growth treatment (annealing of the deposited film in an oxygen atmosphere). Such annealing leads to the saturation of the film with oxygen up to the level required for the formation of an orthorhombic phase.

Methods for synthesizing HTSC films and the corresponding thermal processing in the case of using the ionic sputtering method are described in detail in Ref. 69.

Depending on the temperature of the substrate, the deposited films can be obtained as amorphous, poly- or single crystal. An amorphous film is deposited at a substrate temperature of 500–600 °C.[55] Subsequent annealing at a temperature above the crystallization temperature results in a polycrystalline film. The epitaxial growth of thin single-crystal layers is realized on suitable single crystal substrates when the substrate temperature is higher than the crystallization temperature. Amorphous films, formed on single crystal substrates can become single crystal after annealing at a temperature that is higher than the epitaxy temperature. The saturation of the films with oxygen, depending on their deposition conditions, is realized either in the process of final high-temperature annealing $T > 800$ °C (*ex situ* technology) or (when the sputtering conditions provide a high level of oxygen entry) during step cooling at an oxygen pressure close to 1atm (*in situ* technology). Reference 57 demonstrates the influence of various post-growth processing methods of YBCO films on their superconducting properties.

The above-described post-processing methods of HTSC films provide the necessary superconducting properties and are the most common.

Original methods of post-growth processing of the films, such as atomic oxygen HF plasma treatment,[33] are also used.

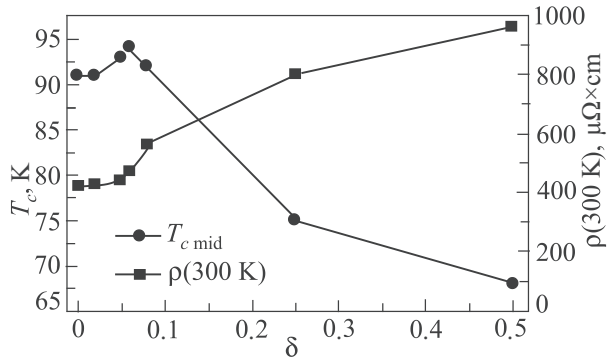

Fig. 6. The transition temperature and resistivity of $YBa_2Cu_3O_{7-\delta}$ film as functions of the oxygen deficit parameter $\delta$.[65]

Figure 6 shows the transition temperature and resistivity of the $YBa_2Cu_3O_{7-\delta}$ films as functions of the degree of oxygen deficit $\delta$.[65] The maximum value of the transition temperature corresponds to the degree of oxygen saturation, equal to 6.94 or $\delta = 0.06$.[58]

### 4.1.4. Chemical film manufacturing methods

Among the mastered chemical methods for manufacturing thin HTSC films the most interesting involves the deposition of thermal decomposition products of highly volatile organometallic precursors onto single crystal substrates (MOCVD, which stands for Metal-Organic Chemical Vapor Deposition). In the MOCVD method the metal components of the film transport the organometallic volatile compounds to the reactor in the form of vapors and mix them with a gaseous oxidant, followed by the decomposition of the vapors in the hot walled reactor or on a heated substrate, and the subsequent formation of a film in the HTSC phase. Metal $\beta$-diketonates are most often used as the volatile compounds.[71] Undoubted advantages of the MOCVD method include, first and foremost, the versatility with respect to the composition of the obtained materials, the possibility of applying homogeneous one- and two-sided films that are homogeneous in composition and thickness to the details of a complex configuration and a large area, including continuous deposition of the film onto extended metal carriers such as tapes.[34,45,58] The structure of one of such tapes, produced by Super Power USA, is shown in Fig. 7.[56] However, there are a number of problems inherent to the MOCVD method. The process of deposition from the gas phase has an incongruent nature and depends on a number of factors: the temperature, total pressure, partial pressures of oxygen, carbon dioxide, and water (oxidation products of the organic component of the utilized compounds), flow rates and their distribution in the reactor and above the substrate, as well as the general composition and homogeneity of the mix of volatile components in the gas phase, etc. Another technological problem is associated with the need to create an optimal morphology of the film. An important problem is the search for substances that have a high and reproducible volatility.[78–81] The Chemical Solution Deposition (CSD) method is currently the most developed and widely used for manufacturing buffer oxide and HTSC films.[59,61,82] The application of precursors to substrates in this method is carried out using salt solutions[83] or the sol-gel process.[60] Currently the CSD method involving precursors based on trifluoroacetates (TFA)[84–86] is successfully used to synthesize $YBa_2Cu_3O_{7-\delta}$ films. In this method the precursor solution is obtained by dissolving yttrium, barium, and copper trifluoroacetates in methanol. The precursor deposited onto the substrate is subjected to further two-stage heat treatment to convert the precursor layer into the $YBa_2Cu_3O_{7-\delta}$ film. In the first stage (pyrolysis) the rate of the process is limited to very low heating speeds, due to the development of shrinkage stress in the films.[35]

Considerable efforts are made to replace TFA precursors with substances that do not contain fluorine, and to also set water and oxygen pressures during YBCO film synthesis in order to increase their growth rate. An important advantage of this technology is the possibility of including rare-earth elements and zirconium oxide in the precursors in order to create artificial pinning centers in the films.[35,36]

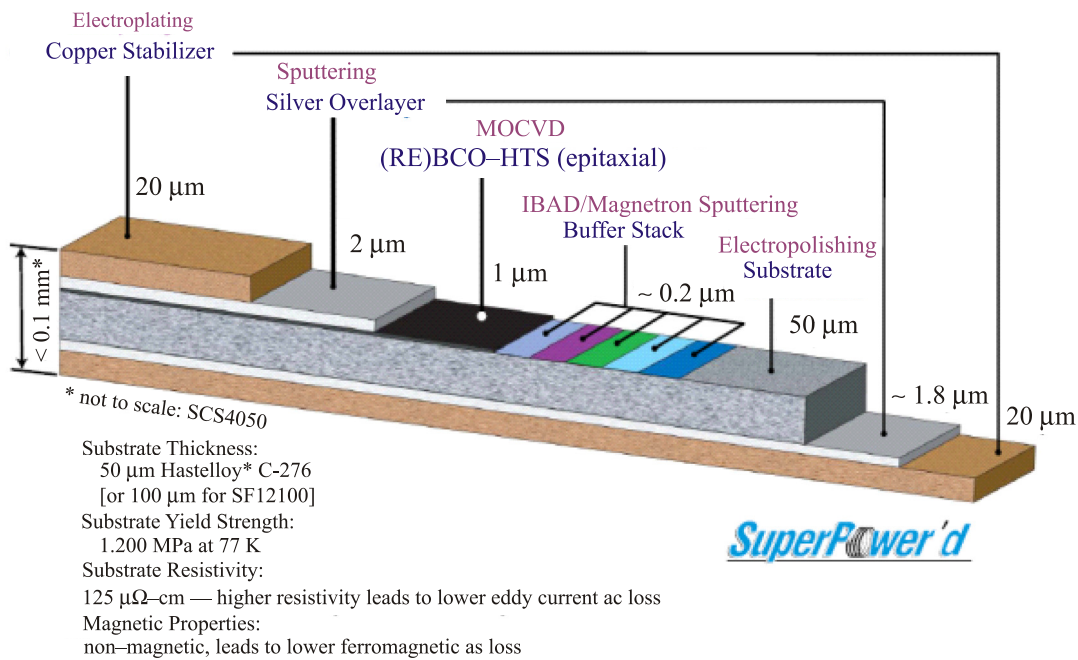

Fig. 7. A diagram of how a HTSC wire in the shape of a ribbon, based on $YBa_2Cu_3O_{7-\delta}$ film with a thickness of 1 $\mu m$, is constructed by Super Power USA.[56]

### 4.2. Bulk samples

#### 4.2.1. Polycrystalline ceramics

Ceramic solid-phase synthesis, similar to the method used by Bednorz and Müller in the discovery of the lanthanum-based compound,[1,87] was the method used to obtain the first bulk samples of $YBa_2Cu_3O_{7-\delta}$ tablets. It involves mixing, annealing, and sintering the powder components. The powders are precursors taken in stoichiometric ratio from oxides and carbonates ($Y_2O_3$, $BaCO_3$, and $CuO$), and ground and mixed in a porcelain mortar or in a ball mill. The mixture of powders is annealed at a temperature of $800\,°C–950\,°C$ for 8–24 h. The mixture is then cooled, crushed, and fired. This process is repeated several times in order to obtain a uniform material. The mixture is pressed into tablets under a pressure of $(0.7–7.2) \times 10^8$ Pa and sintered. The parameters of the sintering process, such as temperature, annealing time, atmosphere and cooling rate, play a very important role in obtaining high quality HTSC materials. If the powders are annealed at a temperature of $800–950\,°C$, then the tablets are sintered at a temperature of $950\,°C$ in an oxygen atmosphere for 6–12 h. The tablets are then slowly cooled and annealed in an oxygen atmosphere at $425–500\,°C$ for 6–12 h. Then the tablets are cooled to room temperature at a rate of $50\,°C/h$. The stoichiometry of oxygen in this material is very important for obtaining a superconducting $YBa_2Cu_3O_{7-\delta}$ compound. During sintering, a $YBa_2Cu_3O_6$ compound with a tetragonal structure is formed and during slow cooling in an oxygen atmosphere it turns into a superconducting $YBa_2Cu_3O_{7-\delta}$ with an orthorhombic lattice structure. The absorption and loss of oxygen are reversible in $YBa_2Cu_3O_{7-\delta}$. A fully oxidized orthorhombic $YBa_2Cu_3O_{7-\delta}$ sample can be converted into a tetragonal $YBa_2Cu_3O_6$ by heating in vacuum at a temperature above $700\,°C$. A disadvantage of the method is its duration, caused by the large grain sizes (granules) of the ceramic material and the inhomogeneity of the reagent mixture. The growth of the crystallites is not controlled, and therefore the electrical and magnetic properties of the ceramics are not reproduced. The chemical methods for obtaining the powders (sol-gel method, spray drying, aerosol pyrolysis, coprecipitation) yield more uniform ceramic samples.[88]

#### 4.2.2. Methods for obtaining single crystals

Let us now consider the methods involved in obtaining REBCO (RE = Y, Nd, Sm, Pr) single crystals. The goal of single crystal technology, as a rule, is to grow a chemically and structurally uniform crystal of a given size, shape, and chemical composition with a low level of defects and impurities. There are basic methods for growing HTSC single crystals from melts:[25] spontaneous crystallization of melt solutions, the Bridgman method, zone melting method, Kyropoulos and Czochralski methods (the method of drawing out the seed crystal from a supercooled melt). The most common mechanisms of crystallization and growth of crystal faces from among the listed methods is growth according to the Frank-Cabrera mechanism, as well as layered growth.

The crystals obtained using the spontaneous crystallization method[89–91] usually represent thick plates with the smallest thickness along the direction $\langle 001 \rangle$, since at high cooling rates the growth rate of $\{100\}$ faces is approximately 5 times higher than the growth rate of $\{001\}$ faces. If we reduce the cooling rate to $(0.5–1)\,°C/h$, then thick parallelepiped prisms and isometric crystals are formed.

A special place in the family of REBCO single crystals is occupied by PrBCO,[92] which is not a superconductor when prepared according to the methods mentioned above. When using the floating zone method it is possible to obtain superconducting PrBCO crystals[92] when oxygen partial pressure is reduced. This is probably due to the noticeable change in the structural parameters of the single crystal, which in turn leads to a screening of the Pr magnetic moment.[92]

When it comes to the method of drawing the seed crystal from a supercooled melt[92–97] low supersaturations are normal, and the hydrodynamic conditions for the distribution of temperatures and concentrations of the nearby growing crystal play the main role in crystal growth. When controlling the drawing rate and the growth anisotropy, it is possible to obtain pyramidal crystals with an expanding base, large isometric crystals with different beveled edges, pyramidal crystals with a "concave" bottom face, and cylindrical crystals.

A modified Czochralski method[92,93] (SRL-CP: Solute Rich Liquid-Crystal Pulling, TSSG: Top Seeded Solution Growth) allows to grow large single crystals (Fig. 8) of almost any REBCO phase (RE = Y, Nd, Sm) as well as $SmYBa_2Cu_3O_{7-\delta}$ and $NdYBa_2Cu_3O_{7-\delta}$ solid solutions.

In the continuous crystal growth method[98] the growing single crystal is replenished under steady-state (stationary) conditions by separating the zones of crystal growth ($T < T_p$) and phase dissolution $Y_2BaCuO_5$ ($T > T_p$, $T_p$ is the peritectic temperature) by using the melt layer. In case of a temperature gradient (about 3 °C/cm), diffusion and convection transport of a melt saturated with yttrium occurs through this layer. Yttrium oxide crucibles are used in order to exclude crystal contamination with crucible material.

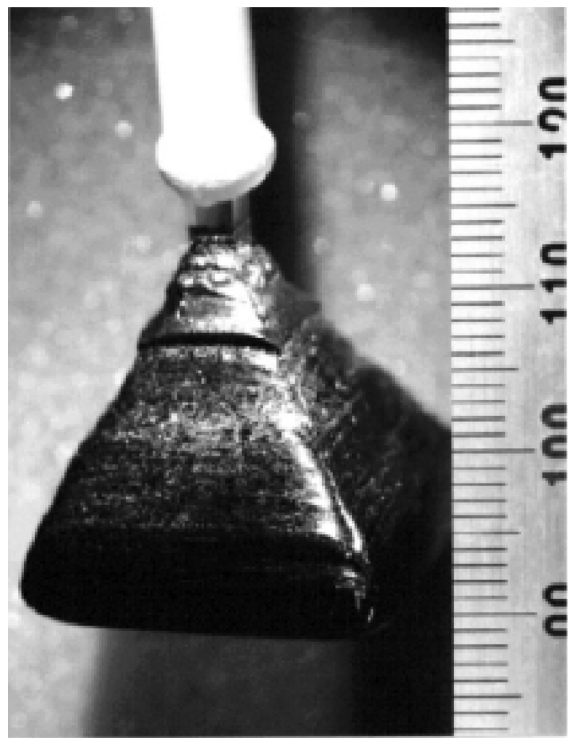

Fig. 8. A large $NdBa_2Cu_3O_{7-\delta}$ single crystal, measuring $24 \times 24$ in the $ab$ plane and 21 mm in the direction of the $c$ axis.[93]

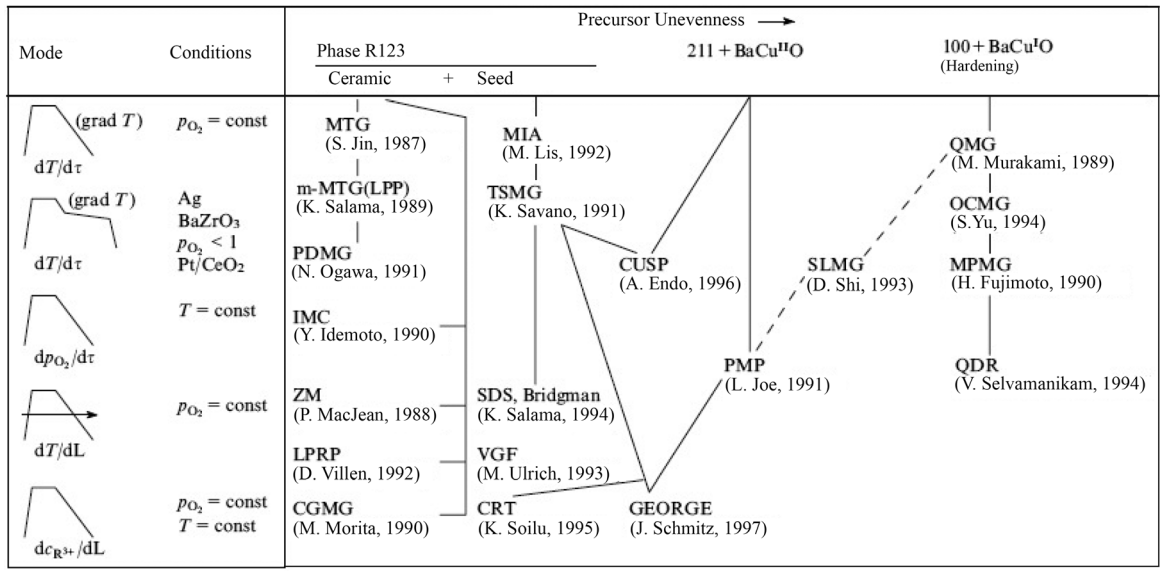

Fig. 9. Melt methods for producing coarse-grained ceramics.[13]

Recently methods for obtaining single crystal HTSC whiskers[99,100] have been in development, since similar crystals can have unique electrophysical and mechanical properties.

We will now transition to methods of manufacturing textured large quasi-single crystals. By virtue of their layered structure all HTSCs under consideration have a crystallographic anisotropy of their physical properties, which leads to the need to create a certain ordered structure (texture) of the polycrystalline material. The main microstructural motif of coarse-grained ceramics is an ensemble of large (0.5–10 cm) pseudomonocrystalline formations that are separated by high-angle boundaries. Each of these formations is a packet of thin (5–50 $\mu$m) $YBa_2Cu_3O_{7-\delta}$ plates or lamellae with a length-to-thickness ratio of about 1000. The plates arranged in parallel to each other are separated by small-angle boundaries that are "transparent" for the critical current.

There are about 20 different types of melt methods for producing HTSCs, considering the precursor chemical type, mechanical history, and heat treatment regime (Fig. 9).[13] One of the first was the MTG (Melt Textured Growth)[101] method, which was modified into the LPP (Liquid Phase Processing) method.[102] These methods can be divided[13] into: (1) methods for increasing the dispersity and homogeneity of the secondary-phase particle distribution, such as QMG (Quenched Melt Growth),[103] PDMG (Platinum Doped Melt Growth),[104,105] MPMG (Melt Powder Melt Growth),[106] PMP (Powder Melt Process),[107] SLMG (Solid Liquid Melt Growth),[108] QDR (Quench and Directional Recrystallization),[109] LPRP (Liquid Phase Removal Process);[110] (2) a method based on the influence of a gas atmosphere, known as IMC (Isothermal Melt Crystallization);[111–113] (3) methods of forming a single-domain structure, such as VGF (Vertical Gradient Freeze), SDS (Seeded Directional Solidification), ZM (Zone Melt), TSMG (Top Seeded Melt Growth),[114] GEORGE (Geometrically Organized Growth Evaluation), CGMG

(Constitutional Gradient Melt Growth),[115] MIA (Magnetically Induced Alignment),[116,117] CUSP (Constant Undercooling Solidification Processing),[118] CRT (Composite Reaction Texturing);[119] (4) the method of chemically modifying and creating effective pinning centers, called OCMG (Oxygen Controlled Melt Growth).[120,121] In our opinion, methods referred to as TSMG and OCMG are of greatest interest when it comes to practical HTSC application. In the TSMG method artificial formation and growth centers of the REBCO phase[114] are formed with the introduction of single, relatively large primers from rare-earth analogues of the REBCO phase having a higher peritectic decay temperature (for SmBCO $T_p \approx 1050$ °C, for NdBCO $T_p \approx 1080$ °C). The seed crystal is placed on the upper part of the dense preform (Fig. 10) and a melting-crystallization cycle is conducted, using a temperature gradient or by moving the high-temperature zone above the sample. The seed initiates the formation of the main phase along the crystallization front, which leads to the appearance of gigantic pseudomonocrystalline domains, the dimensions of which are comparable to the size of the sample itself (Fig. 11). At the same time, the orientation of the resulting pseudomonocrystals is practically the same as the orientation of the seed crystal.

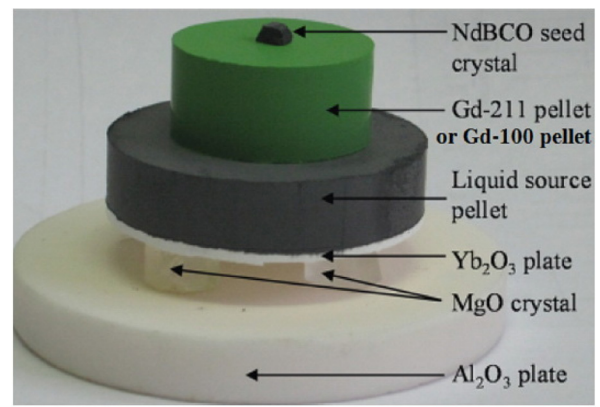

Fig. 10. A model image of the substances and their arrangement with respect to one another during the growth of a GdBCO single crystal using the Czochralski method with a seed in the form of a NdBCO single crystal.[122]

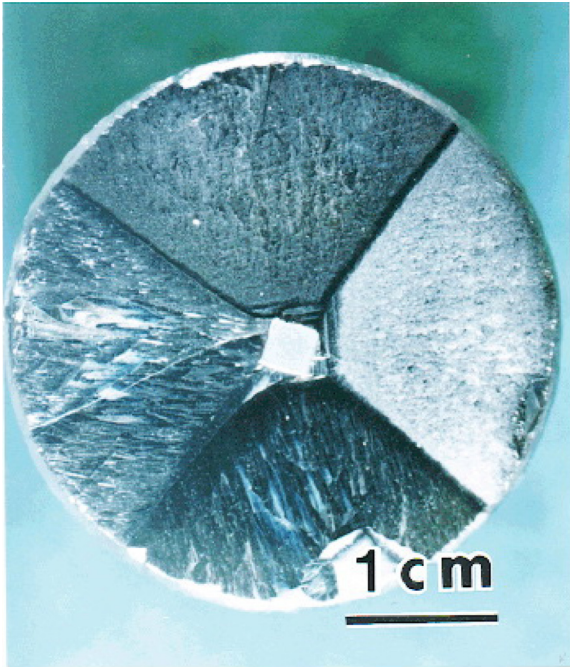

Fig. 11. The external appearance of a SmBa$_2$Cu$_3$O$_{7-\delta}$ quasi-single crystal disc with four characteristic growth sectors and the boundaries of these sectors.[121]

In the OCMG method[120,121] the HTSC materials are obtained at a reduced oxygen partial pressure (0.1–1 mol. % O$_2$). The method is based on the fact that rare-earth elements with the largest ion radii (Nd, Sm, Eu, Gd) are able to form compounds such as RE$_{1+x}$Ba$_{2-x}$Cu$_3$O$_{7-x}$. When crystallized from the melt at a reduced oxygen partial pressure the degree of barium substitution is significantly less and the superconducting transition temperature increases significantly (up to 95–96 K). This fact is also associated with the possible cation ordering in the crystal lattice, such as the formation of neodymium ion pairs in barium positions, which leads to a decrease in the oxygen sublattice disorder. On the other hand, this type of a superconducting matrix could have areas with fluctuating chemical composition, which are effective pinning centers. In a nonzero magnetic field their superconductivity is sharply suppressed at the liquid nitrogen boiling point, resulting in the peak effect (see Sec. 5.2.1).

To date, the record sizes of textured samples are single-domain GdBaCuO discs with a diameter of 143–153 mm and a thickness of 20 mm.[123–126] In Ref. 123 a GdBaCuO disc was grown using the composite gradient method (the concentration of Dy increases and the concentration of Gd decreases as we move from the center to the periphery of the disc).

## 5. The main superconducting properties of (RE)Ba$_2$Cu$_3$O$_{7-\delta}$

Based on studies involving LTSCs it is known that superconducting alloys and chemical compounds are type-II superconductors.[127,128] (RE)Ba$_2$Cu$_3$O$_{7-\delta}$ compounds belong to this group. The first $H_{c1}$ and second $H_{c2}$ critical fields, as well as the critical current density $j_c$ are the most important parameters from the point of view of type-II superconductor application, and as a rule they are determined experimentally.

### 5.1. The magnitude and anisotropy of the critical field, coherence length, and magnetic field penetration depth into the superconductor

The theory of isotropic type-II superconductors[129] tells us that

$$H_{c1} = \frac{\Phi_0}{4\pi\lambda^2}\ln\kappa, \qquad (2)$$

where $\Phi_0$ is the magnetic flux quantum, $\lambda$ is magnetic field penetration depth in the superconductor, $\kappa = \lambda/\xi$ is the Ginzburg-Landau parameter, and $\xi$ is the coherence length. For type-II superconductors $\kappa > 1/\sqrt{2}$.[130] Respectively,

$$H_{c2} = \frac{\Phi_0}{2\pi\xi^2}. \qquad (3)$$

The values of $\xi$, $\lambda$ and $\kappa$ can be calculated based on Eqs. (2) and (3) while accounting for the relation for $\kappa$. In contrast to isotropic low temperature superconductors, the following relationships between the critical fields in certain mutually perpendicular directions and the parameters of penetration depth and coherence length were obtained in Ref. 129 within the framework of the Ginzburg-Landau (GL) phenomenological theory of superconductivity for the case of layered anisotropic superconductors, such as the HTSCs we are considering:

$$H_{c1}^c = \frac{\Phi_0}{4\pi\lambda_{ab}^2}(\ln\kappa_c + 0.5), \qquad (4)$$

$$H_{c1}^c = \frac{\Phi_0}{4\pi\lambda_{ab}\lambda_c}(\ln\kappa_{ab} + 0.5), \qquad (5)$$

$$H_{c2}^c = \frac{\Phi_0}{2\pi\xi_{ab}^2}, \qquad (6)$$

$$H_{c2}^{ab} = \frac{\Phi_0}{2\pi\xi_{ab}\xi_c}, \qquad (7)$$

$$\frac{m_c}{m_{ab}} = \frac{\Phi_0}{2\pi\xi_{ab}\xi_c}, \qquad (8)$$

wherein the subscripts $c$ and $ab$ denote the direction of the magnetic field along the $c$ axis of the crystal and along the $ab$ plane of the crystal $\kappa_c = \lambda_{ab}/\xi_{ab}$, $\kappa_{ab} = \lambda_c/\xi_{ab}$. In this case

$$\frac{\lambda_c}{\lambda_{ab}} = \frac{\xi_{ab}}{\xi_c} = \left(\frac{m_c}{m_{ab}}\right)^{0.5}, \qquad (9)$$

where $m_c$ and $m_{ab}$ are the effective electron masses along the indicated directions. Considering that in the given HTSCs $m_c/m_{ab} \gg 1$ we get

$$\frac{H_{c2}^{ab}}{H_{c2}^c} = \frac{H_{c1}^c}{H_{c1}^{ab}}. \qquad (10)$$

Thus, to determine the parameters $\lambda_c$, $\lambda_{ab}$, $\xi_{ab}$, $\xi_c$ it is necessary to produce the measurements for $H_{c1}^c$, $H_{c1}^{ab}$, $H_{c2}^c$, and $H_{c2}^{ab}$. Single crystal samples are most often used for the indicated measurements. According to the GL theory, the temperature dependence of $H_{c1}$ is described by equation

$$H_{c1}(T) = H_c(0)\left[1 - \left(\frac{T}{T_c}\right)\right]^2, \qquad (11)$$

where $H_c(0)$ is the thermodynamic critical field. Equations (2)–(11) can be used as qualitative estimates for HTSCs.

The first critical field, its temperature dependence, and the anisotropy of the YBa$_2$Cu$_3$O$_{7-\delta}$ compound were first measured on single crystal samples, and then on polycrystalline solids. At the same time, parameters such as $\xi$ and $\lambda$ were determined with the help of these measurements. In the case of single crystals the $H_{c1}$ field was measured at two directions of the external field $\mathbf{H}$, directed along the $c$ axis of the crystal and across this axis. Usually different variations of the method for measuring the sample's temperature dependence of magnetization (magnetic moment) in different magnetic fields are used in order to determine $H_{c1}$. Figure 12(a) shows the experimental dependences of $H_{c1}(T)$ obtained from one of the earlier studies, Ref. 131. Figure 12(b) shows the dependences $H_{c1}(T)$ obtained later using a more valid version of this method.[132] Figure 12(c) shows the dependence $H_{c1}^{ab}(T)$, obtained using the data on high-frequency (6 MHz) measurements of $\lambda$ for a YBa$_2$Cu$_3$O$_{7-\delta}$ single crystal.[133] In particular, Fig. 12(c) shows that the temperature dependence is practically the same as the curve calculated using Eq. (11) (the solid curve on the figure is the dependence calculated according to the BCS theory).

In the case of polycrystalline materials the problem of determining $H_{c1}$ and the anisotropic parameters $m_c$ and $m_{ab}$ for YBa$_2$Cu$_3$O$_{6.95}$ granules was solved in Refs. 134–136. The method of determining $H_{c1}(77\,\mathrm{K})$ is based on the measurements of the critical current of a flat textured ceramic sample as a function of the magnetic field trapped at different angles relative to the sample. In this case the dependences $H_{c1}(T)$ are close to those obtained for single crystals.

We will now turn to studies pertaining to defining $H_{c2}^{ab}$. The upper critical field is determined according to the following methods:[129] the resistive method, the fluctuation conductivity method, the reversible magnetization method, the pinning method, the heat capacity method, and the optical method. The temperature dependence $H_{c2}^{ab}(T)$ is given at $T - T_c \ll T_c$ by expression[129]

$$H_{c2}^{ab}(T) \sim (T_c - T)^{0.5}. \tag{12}$$

Figure 13 shows the first ever dependences $H_{c2}^{ab}(T)$ and $H_{c2}^{c}(T)$ obtained in a broad range of temperatures, all the way to 4 K.[137] The authors of the study believe that these dependences are consistent with the WHH theory.[138,139] According to this theory, the experimental dependences $H_{c2}^{ab}(T)$ prove that their explanation requires the consideration of spin-Zeeman ($\alpha$) and the spin-orbit ($\lambda_{SO}$) effects caused by the magnetic field's impact on the superconductor. As far as the dependence $H_{c2}^c(T)$ is concerned, it can be explained without taking into account these effects ($\alpha$ and $\lambda_{SO}$ are equal to zero).

### 5.2. Vortices, pinning of vortices, critical current density

The ceramic technology for obtaining bulk HTSC samples either via solid-state reaction during the sintering of oxides of different metals[25] or by applying the melt method to the oxides and salts of these metals[13] yields superconductors that vary in density and morphology. In general, two types of ceramics are known: granular and single crystal (or pseudomonocrystal). The first are a set of single crystal microgranules of an HTSC having an irregular shape with typical dimensions from one to 20 $\mu$m with a random arrangement of the crystallographic $c$ axis in space. Depending on the features of the technology used to prepare the granular samples (discs, cylinders, etc.) the density of the HTSC material can vary from 3 to 4.5 g/cm$^3$. As a result, the HTSC sample is a set of single crystals that have been pressed together with different force and separated by weak superconducting bonds, mainly in the form of superconductor-normal metal-superconductor (S-N-S contact) Josephson junctions. The critical temperature of the granular samples is determined by the granule properties, and the transport properties are defined by the weak superconductivity of the Josephson junctions. This kind of ceramic is often called a natural Josephson medium. These cuprate HTSCs are referred to as a first-generation high-temperature superconductors. It is with the help of this type of ceramic that the discovery of high-temperature superconductivity with $T_c = 93$ K in the Y-Ba-Cu-O$^2$ compound was made. The critical current density at $T = 77$ K in granular HTSCs is very small ($10^2$–$10^3$ A/cm$^2$). Second-generation cuprate HTSCs, the superconducting transition temperature of which is close to the value above, have almost no weak S-N-S junctions, and are characterized by significantly higher values of critical current density ($\sim10^5$ A/cm$^2$) and have a high density (up to 5.7 g/cm$^3$). The technology involved in manufacturing this type of HTSC is described in Sec. 4.2.2.

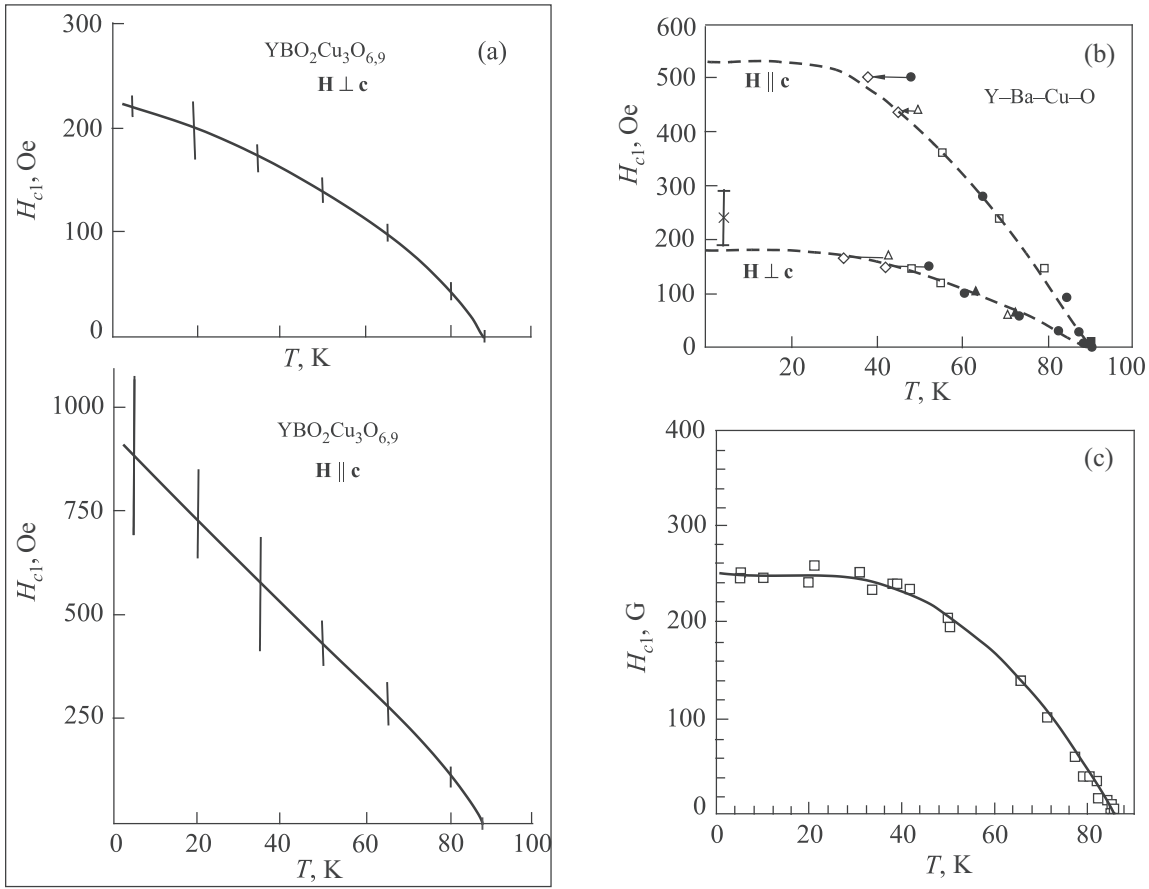

Fig. 12. The temperature dependences of the first critical field $H_{c1}$ for YBa$_2$Cu$_3$O$_{7-\delta}$ single crystals, obtained using various methods with a slight difference in oxygen content for two field directions relative to the $c$ axis: YBa$_2$Cu$_3$O$_{6.9}$ crystals[131] (a), crystals with $T_c = 90$ K (Ref. 132) (b), crystals with $T_c = 86.2$ K for a direction of the field perpendicular to the $c$ axis, the curve corresponds to the calculation performed in accordance with BCS theory[133] (c).

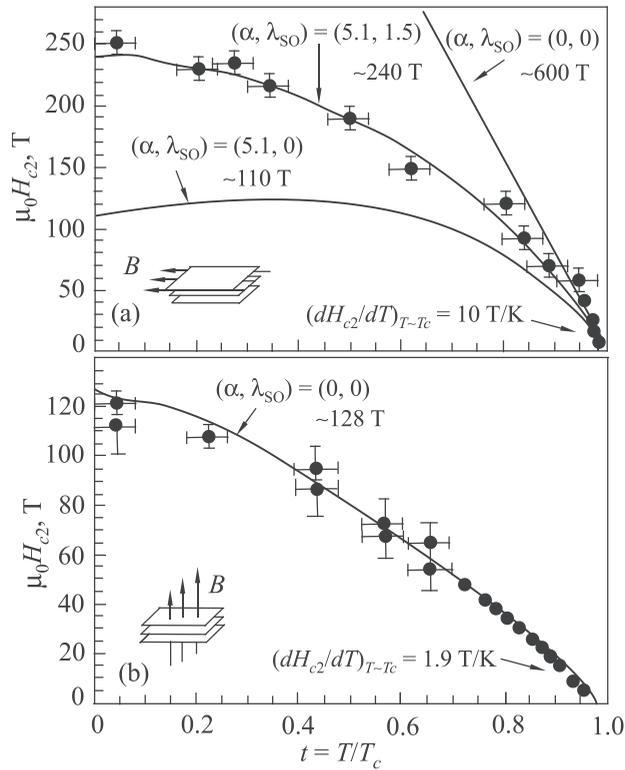

Fig. 13. Experimental points of the temperature dependences of the second critical field of a YBa$_2$Cu$_3$O$_{7-\delta}$ single crystal along and across the $ab$ plane, the solid curves are the WHH theory (1 T = $10^4$ Oe = $10^4$ G).[137]

Different types of HTSC ceramics are distinguished according to the features of their magnetic and transport properties. In particular, this is manifested by the different vortex structures of these HTSCs in a magnetic field exceeding $H_{c1}$, and their difference from the vortex structure of a type-II LTSC. First we will discuss the vortex structures of granular ceramics. If only one type of current vortex is known in a mixed-state LTSC, especially if this is the Abrikosov vortex,[140] then at least two types of vortices can exist in a granular HTSC. In addition to Abrikosov vortices, these are so-called Josephson vortices.[141] In accordance with this, different critical magnetic fields can be observed, starting from the weakest $H_{c1J} \approx 1$–10 Oe, which corresponds to the onset of magnetic field penetration in the form of Josephson vortices into the intergranular interlayers (weak bonds), and then an average $H_{c2J} \approx 10$–100 Oe, which corresponds to the suppression of weak bond superconductivity by a magnetic field, and then a field that is close to the first critical field of the granules ($H_{c1g}$), which indicates the onset of field penetration into the granules in the form of Abrikosov vortices. Complete suppression of HTSC superconductivity occurs once the field reaches the value of the second critical field of the granules ($H_{c2g}$). The indicated values of $H_{c1J}$ correspond to a low critical current density $j_c$ of such HTSCs that is characteristic for Josephson junctions ($j_c \approx 10^2$–$10^3$ A/cm$^2$). This feature of HTSCs limits the use of granular ceramics by low-current devices (SQUIDs, weak field magnetic screens, etc.). However, granular ceramics are convenient for modeling processes associated with trapping the magnetic field in superconductors (see Sec. 5.3.1) since in this case no complex or expensive sources of a strong magnetic field are required.

### 5.2.1. Properties of second-generation HTSC ceramics in a mixed state

The practical problems of using HTSCs require an increase in the density of the pinning center density of the vortices and their activation energy $U$ in order to obtain high values of $j_c$, but with conservation of the HTSC high critical temperature.[142] An increase in $j_c$ is presently achieved by technological efforts.

When the value of the external magnetic field $H \geq H_{c1}$ the field starts to penetrate the HTSC sample as Abrikosov vortices. In this respect, the picture of the mixed state is qualitatively similar to that of LTSC samples. Such a HTSC also has two critical fields $H_{c1}$ and $H_{c2}$. Their peculiarity, as mentioned above, is expressed in the anisotropy of these fields and also by the fact that the lattice of their vortices is square rather than triangular, as is the case for type-II LTSCs. Similar to LTSCs the critical current density through the HTSC sample is determined by the vortex pinning force $f_p$, the external magnetic field $H$ that acts on the sample, and the sample temperature $T$. In accordance to the type-II superconductivity theory[143] the critical current for a given operating temperature $T < T_c$ is determined by the pinning (catching) of the Abrikosov vortices on the microscopic defects of the sample. In an ideal single crystal superconductor without pinning centers the critical current must be equal to zero. In a real superconductor that has pinning centers, when the transport current $I$ reaches the critical value of current density $j_c$ it means that the Lorentz force overcame the pinning force, and the vortices start to move across the direction of the current. In this case voltage appears across the sample, i.e., an electric field $E$ is generated in the direction of the current. The sample goes into a resistive state and energy dissipation occurs. The physical concepts associated with the resistive state in such HTSC materials are close to those of LTSCs.[144] At the same time the HTSC mixed state has singularities. These are determined mainly by the smaller values of $\xi$ (see below) and HTSC anisotropy. The smallness of the $\xi$ value indicates that in HTSCs even nanoscale structural defects can impact the superconducting properties of the samples, especially $j_c$. The complex structure of real HTSC crystals, which often contains different phases, impurities, and lattice distortions, contributes to an increase in the number and the extent of the pinning centers.

The anisotropy of HTSC crystals causes the corresponding anisotropy of $j_c$. It has one very important feature. It turned out that at the stage of $(RE)Ba_2Cu_3O_{7-\delta}$ crystal growth, mechanical stresses arise in the crystal during the phase transition from the tetragonal to the orthorhombic structure of the lattice, which lead to the formation of twinning planes that are parallel to the $c$ axis of the crystal. It has been experimentally determined that the current through the crystal has different critical values depending on the angle between the twinning planes and the direction of the current. This feature of HTSC crystals was confirmed in a number of experimental studies and found its explanation in Refs. 142 and 143. In this way the twinning planes in $YBa_2Cu_3O_{7-\delta}$ are a characteristic region of vortex pinning for these compounds, causing the critical values of the current as a function of the given angle to differ by a factor of almost 30.

In order to create HTSC magnets that generate a large constant magnetic field by passing a current through the multi-turn coil, it is necessary to generate a high $j_c$ not only in the external magnetic field of the Earth, but to also prevent its significant decrease at working values of the magnet's field. The operating field values currently amount to 1.5 T–20 T, and will be even higher in the future. This is achieved not only by special constructive solutions, but also by finding such HTSCs in which the action of the magnetic field on the current-carrying properties of the HTSC conductor would be minimal. For example, it was found that replacing yttrium in the $YBa_2Cu_3O_{7-\delta}$ compound with neodymium $(NdBa_2Cu_3O_{7-\delta})$ decreases the influence of the external field on the critical current of a conductor made of this material. At the same time a peak effect is observed, at which the decrease in the critical current density that occurs with an increase in the field changes to an increase and is followed by a slight decrease (Fig. 14).[121]

At the same time it is clear that the field suppression of the critical current cannot be excluded completely, since in this case there is an effective combination of the transport current in each section of the coil turns and the current induced in this region by a magnetic field from other parts of the coil. As such, in the first approximation, without accounting for anisotropy, it can be assumed that the critical current must decrease in accordance to expression

$$I_c(T,H) \approx I_c(T,0)\left[1 - \frac{H}{H_c(T)}\right], \qquad (13)$$

where $I_c(T,0)$ is the critical current at $H = 0$. Figure 15 shows the experimental dependences of the critical current $I_c$ of a $YBa_2Cu_3O_{7-\delta}$ tape on the magnetic induction $B$ for two of its directions: at the top, along the $ab$ plane, and at the bottom, along the $c$ axis of the superconductor at different temperatures.[145]

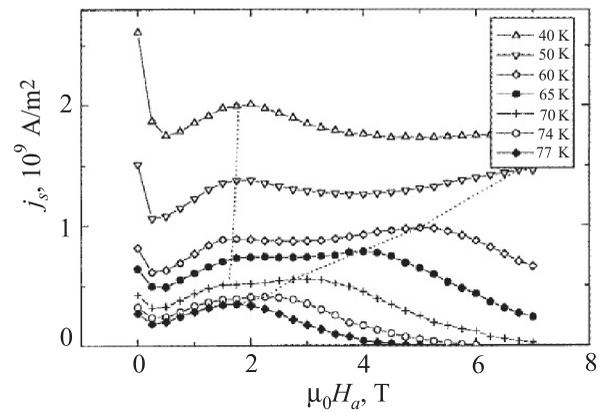

Fig. 14. An illustration of the peak effect on the magneto-field dependence of the critical current density of the $NdBa_2Cu_3O_{7-\delta}$ compound.[121]

The measurements of the critical current density $j_c$ are usually carried out by the four-probe method or by the contactless method with respect to the width $\Delta M$ of the hysteresis loop of the magnetization curve[146]

$$j_c = \frac{3\pi\Delta M}{8rV}, \quad \text{for a cylindrical sample}, \qquad (14)$$

Is. vs. Field Data (Lift Factor Measurement):
SP Wire Id M3-909-3
measurements were performed at the University of Houston

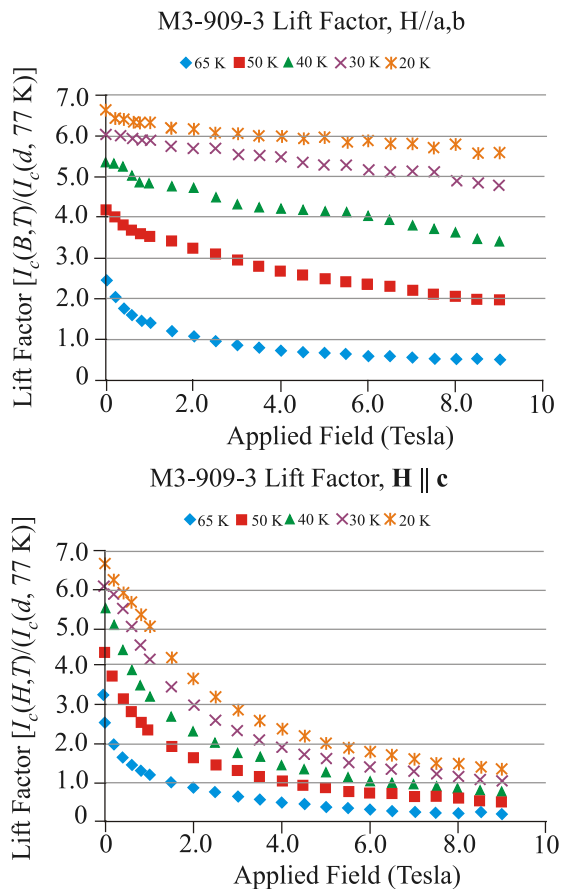

Fig. 15. Magneto-field dependences (at different temperatures and directions of the field) of the current density of a HTSC tape based on $YBa_2Cu_3O_{7-\delta}$ film produced by Super Power.[145]

$$j_c = \frac{3\Delta M}{LV}, \quad \text{for a square thin plate,} \qquad (15)$$

wherein $\Delta M$, $V$, $r$, $L$ is the width of the hysteresis loop, sample volume, cylinder radius, and length of the side of the plate. For YBa$_2$Cu$_3$O$_{7-\delta}$ single crystals we have $j_c \approx 5 \times 10^6$ A/cm$^2$ ($T = 4.2$ K, $H = 0$ Oe) and $j_c \approx (1-5) \times 10^4$ A/cm$^2$ ($T = 77$ K, $H = 0$ Oe). The critical current density in textured ceramics is $j_c = 10^4$–$10^5$ A/cm$^2$ ($T = 77$ K, $H = 0$ Oe). This current density is also characteristic for the granules in granular ceramics ($T = 77$ K, $H = 0$ Oe). In a field of 1 T the critical current density of textured ceramics falls by a factor of 10. In granulated ceramics the critical current density is 10–1000 times smaller than in single crystals and amounts to $j_c = 1$–$10^3$ A/cm$^2$ ($T = 77$ K, $H = 0$ Oe), and in a field of 1000 Oe at $T = 77$ K it falls to $j_c = 0.1$ A/cm$^2$. The causes for these low values of intergranular critical current density $j_{cJ}$ are: a disorientation angle of neighboring grains; the violation of oxygen stoichiometry at the grain boundary regions, and segregation of impurities at the intergranular boundary. The empirical formulas for dependences $j_c(T,B)$ that are more accurate in comparison than Eq. (13) are given below. The temperature dependence of the critical current density for single crystals is

$$j_c(T) = j_c(0)\left[\exp\left(-\frac{T}{T_0}\right) - \exp\left(-\frac{T_c}{T_0}\right)\right], \qquad (16)$$

wherein $T_0 = 10$–$30$ K.

The dependence of the critical current density on the magnetic field for single crystals is given by equation

$$j_c(B) = j_c(0)\exp\left(-\frac{B}{B_0}\right). \qquad (17)$$

For textured ceramics

$$j_c(B) = \frac{j_c(0)}{1 + (B/B_0)^2}, \qquad (18)$$

where $B_0$ is the adjustment coefficient (ranging from tens to hundreds of gauss).

The progress involved in improving the superconducting parameters of superconductors achieved as a result of developing high-quality YBaCuO samples in comparison to LTSC parameters is shown by the phase diagrams in Fig. 16.[145] In particular, the dependence $H_{c2}^{ab}(T)$ was experimentally obtained relatively recently.[121]

### 5.3. A trapped (captured) magnetic field

The ability to freeze (or capture) a magnetic field is another remarkable property of all type-II superconductors, in addition to having zero electrical resistance. Historically the concept of a trapped (captured) magnetic field (TMF) came out of research (Alekseeevsky in the USSR, Schoenberg in England[147]) on magnetization reversal of superconducting rings by an external magnetic field, conducted in the 1930s. In particular it was determined that if the external field keeps increasing at the moment it reaches the critical field of the ring, the field penetrates into the ring. If after this the external field is reduced and at some value (below the critical value) it is turned off, then the field is stored in the ring at the level corresponding to the moment of its shutdown. This is what we mean by a TMF. The physical cause behind this process is similar to the empirical law of Faraday's electromagnetic induction, extended to the case of a closed superconducting circuit in the form of a ring. Researchers also refer to this phenomenon as the law of magnetic flux conservation in a superconducting circuit. In a simply-connected (without holes) and homogeneous type-I and type-II superconductor in an external field below the first critical value, the magnetic field cannot be captured because it is pushed out by the Meissner effect. A TMF is possible in all types of superconductors if they have spatial inhomogeneities in their superconducting parameters (critical current density, critical field) or defects in the form of cracks or caverns. This is especially applicable to type-II superconductors in an external field larger than the first critical field. Each such defect is a section in the form of a doubly connected or multiply connected superconductor. As a result, any transitions from the superconducting state to normal and back in a magnetic field cause the appearance of local and integral TMFs. A TMF can be displaced from the sample by the Lorentz force. This causes the TMF to decay in time. The Lorentz force arises as a result of the interaction between the TMF and the superconducting current that supports it. The TMF decay rate, i.e., its relaxation, depends on the pinning force of the TMF magnetic flux and the TMF magnitude.

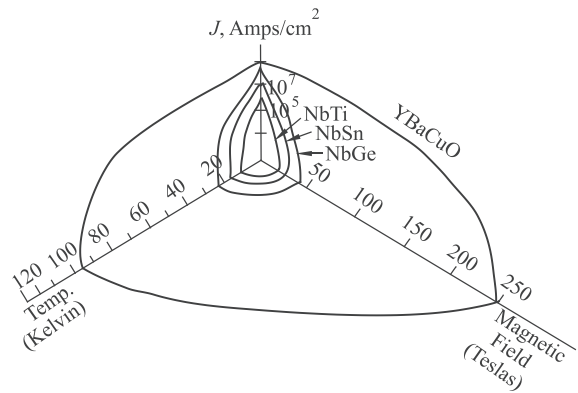

Fig. 16. The physical phase diagrams that generalize the averaged dependences of the critical current density on the magnetic field and temperature, as well as the critical field as a function of temperature, for the YBa$_2$Cu$_3$O$_{7-\delta}$ compound and several types of LTSCs (NbTi, Nb-Sn, Nb-Ge).[145]

The study of the processes responsible for the formation of a trapped magnetic field (TMF) in HTSCs is of great practical importance because currently there is no technology available that can create magnets in the form of short-circuited coils from a long HTSC without consuming current. In this situation a prototype of such an HTSC magnet that does not consume energy could be a bulk HTSC with a trapped field. In addition, capturing a field in a HTSC has purely scientific importance. The process of freezing a field is used, as noted above, in methods for determining the first critical field of the compound YBa$_2$Cu$_3$O$_{7-\delta}$. The study of the TMF in HTSC samples that could be prototypes of HTSC magnets served as the basis of a whole new direction in magnetic studies of HTSCs given an excitation of a TMF using a uniform external field. A TMF changes the transport properties of a HTSC, and carries information about the critical parameters of the HTSC, as well as details about their spatial uniformity.

The first (classical) direction in the formation and study of TMFs is the way a superconducting sample is impacted by external constant magnetic field that is homogeneous with respect to its volume. Reference 148 presents the experimental results that demonstrate clearly for the first time the "evolution" of the trapped magnetic field $YBa_2Cu_3O_{7-\delta}$ in a granular ceramic $YBa_2Cu_3O_{7-\delta}$ over a broad range of an external uniform magnetic excitation field. Figure 17 shows the typical dependence of a homogeneous TMF on an external uniform magnetic field that is perpendicular to the disc surface of a $YBa_2Cu_3O_{7-\delta}$ ceramic (9 mm diameter, 1 mm thick), in a ZFC (Zero Field Cooling) mode.

The figure shows that the critical field at which the magnetic field starts to penetrate the intergranular intervals is $H_{c1J} = 2.5$ Oe, the critical field of total penetration into the intergranular intervals is $H_{tJ} = 30$ Oe, the critical magnetic field of Abrikosov vortice penetration into granules is $H_{c1g} = 50$ Oe, and the critical field of total penetration by Abrikosov vortices into granules is $H_{tg} = 1000$ Oe. The field $H_{tg}$ corresponds to the field at which a critical state is established in the granules, i.e., when the TMF is saturated. The value of $H_{tg}$ ($H_{tg} \ll H_{c2}$) depends on the quantity, dimensions, depths, and relative location of the pinning centers. In the field region of 30–50 Oe (first plateau) a critical state is established across the entire intergranular space. In the field above 50 Oe the concentration of Abrikosov vortices increases in the granules and the scattering fields of these vortices suppress weak bonds among the granules. In an external field $H \geq 1000$ Oe the captured field in the granules reaches saturation, and the TMF value ceases to change (second plateau in Fig. 17). As can be seen from the dynamics, at the initial stage, when the field just starts to penetrate into the sample, in the Josephson medium macroscopic superconducting circuits consisting of granules connected by weak Josephson bonds participate in the TMF. In the low-field region, when the critical state is just getting established in the Josephson medium, the TMF value, which is $H_{tr}$ for highly inhomogeneous samples, can be comparable to the external reference field $H_0$, i.e., 10–30 Oe. Since the size of the contours is much larger than the size of the measuring probe (in this case the Hall converter, HC) the TMF can be easily recorded. Starting from a field comparable with the first magnetic field of the granules $H_{c1g} \sim 50$–60 Oe, the field penetrates into the granules. However, due to the fact that the $H_{tr}$ force lines close inside the sample (around the sample's "own" granules) instead of around the sample, and the pellet size is much less than the size of the HC working area (0.15 × 0.45 mm), the average $H_{tr}$ at $H_0 = 1000$ Oe measured by the HC was ~40 Oe. As shown in the next article, the intra-subcrystallite currents for cuprate HTSCs at $H_0 = 1000$ Oe are about an order of magnitude larger than the Ginzburg-Landau pair-breaking current at $H_0 = 0$ Oe.[149] Consequently the levels of local $H_{tr}$ depending on the depths of the pinning centers and their mutual arrangement can reach the level of $H_{c2} \sim 50$–100 T. The critical density of the intragranular current for a given current with an average pellet size of 10 μm at $T = 77$ K is $j_{cg} = 10^5$ A/cm$^2$. The critical density of the intragranular current in this study was determined using the effective magnetization current $I_H = cMt$ (where in CGSM units: $c$ is the speed of light, $M$ is the density of the sample's magnetic moment, and $t$ is the thickness of the disc). The critical density of intergranular current ($j_{cJ}$ determines the limiting current-carrying capacity of the sample) is found using expression

$$j_{cJ} = \frac{10}{4\pi} \frac{H_{fm}}{t \operatorname{arcsh}\left(\dfrac{R}{2t}\right)}, \qquad (19)$$

where $H_{fm}$ is the maximum TMF value at the section of the first plateau, $R$ is the disc radius, and $t$ is the disc thickness.

The estimate based on Eq. (19) yielded a value $j_{cJ} = 40$ A/cm$^2$ and agrees with both the value estimated according to the Bean model[150] for the field of complete penetration into the center of the disc (22.6 Oe), and with the value measured using the standard four-probe technique. An estimate based on the formula $\mathbf{f}_P = \mathbf{j}_c \times \Phi_0/10$ was given in Ref. 148 for the pinning force of the Abrikosov vortex inside the granule $f_{Pg} = 2 \times 10^{-3}$ dyne/cm and for the vortex in the intergranular space $f_{PJ} = 10^{-7}$ dyne/cm. First of all, these results confirm the correctness of Eq. (19), which ties the critical current density with the value of the trapped field and the sample dimensions, and second, they allow us to determine both the critical current density and the average pinning force via a direct non-contact method.

The study of the TMF spatial distribution can serve as an express method for diagnosing and controlling the quality of HTSC materials.

The measured values of the first critical magnetic fields (they were determined based on the values of $H_0$ at which the magneto-field dependence of the sample's effective demagnetizing factor $n_{eff}$ ($H_0$) has singularities) are used to study the impact crystallite demagnetization fields have on the magnitude of intercrystallite and intracrystallite critical currents, and the values of critical current densities of a quasi-single crystal sample ($5.7 \times 10^3$ A/cm$^2$), crystallites, and subcrystallites ($8.2 \times 10^7$ A/cm$^2$) at $T = 77.4$ K. It is shown that an increase in the size of the crystallites leads to the suppression of their critical current at lower fields, i.e., in crystallites with a small demagnetizing factor the superconductivity is suppressed less than in larger crystallites.[149]

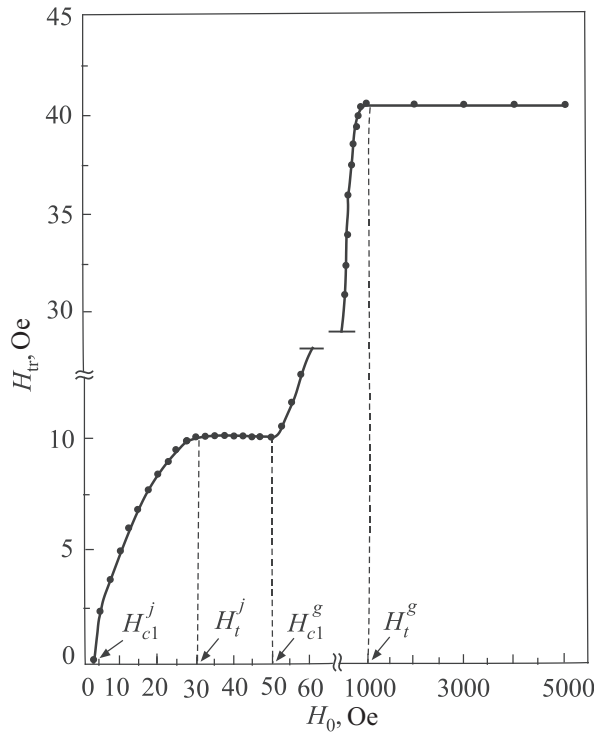

Fig. 17. The dependence of the TMF value in a ceramic $YBa_2Cu_3O_{7-\delta}$ disc on the momentum of the external homogeneous magnetic field at $T = 77$ K.[148]

There are three main modes of magnetic flux trapping in bulk HTSCs:[151,152] Field Cooling (FC), when the sample is cooled in a magnetic field, Zero Field Cooling (ZFC), when it is cooled at zero magnetic field and then the field is turned on and off, and Pulsed Field Magnetization (PFM), when the sample is cooled in a zero magnetic field and is then subjected to an external pulsed magnetic field. In turn, variants of the PFM mode include the Modified Multi Pulse technique combined with Stepwise Cooling (MMPSC),[152] Iterative pulse field Magnetization method with Reduced Amplitude (IMRA),[154–156] Sequential Pulsed Application (SPA).[157] All of these modes are implemented under the influence of a homogeneous excitation field.

The PFM method is compact, inexpensive, and mobile, however it does not give large trapped magnetic fields because the sample is heated when the pulsed field is applied to it. After onset the TMF decreases significantly (relaxes) in a time of 100–300 ms and is always less than in the FC regime (Fig. 18).[154] The task of increasing the relaxation time is now one of the most urgent. The record value of TMF $B_{tr}$ in the PMF-MMPSC mode is 5.2 T at $T = 29$ K, obtained on a GdBaCuO sample with a diameter of 45 mm and thickness of 15 mm.[153]

The record values of $B_{tr}$ in the FC mode are $B_{tr} = 17.24$ T at $T = 29$ K,[158] in the gap between two YBaCuO discs with a diameter of 26.5 mm and a thickness of 15 mm each (Fig. 19), as well as the recently obtained value $B_{tr} = 17.6$ T at $T = 26$ K (Ref. 159) in the gap between two GdBaCuO discs with a diameter of 24.15 mm and thickness of 13 mm each.

It should be noted that a typical TMF value in the FC mode for YBaCuO discs with a diameter of 20–60 mm at the boiling point of liquid nitrogen is $B_{tr} = 1$–1.5 T (Fig. 19), and the magnitude of the critical current density $j_c = 10^4$–$10^5$ A/cm$^2$ ($H = 0$), which decreases to 5–30 kA/cm$^2$ in a 1 T field. Doping the sample with zinc[160,161] creates effective pinning centers, and adding platinum and uranium to the HTSC preparation process, in combination with irradiation with warm neutrons, leads to an increase in $B_{tr}$ from 0.65 T to 2.04 T at a temperature of 77 K (Fig. 20).[162]

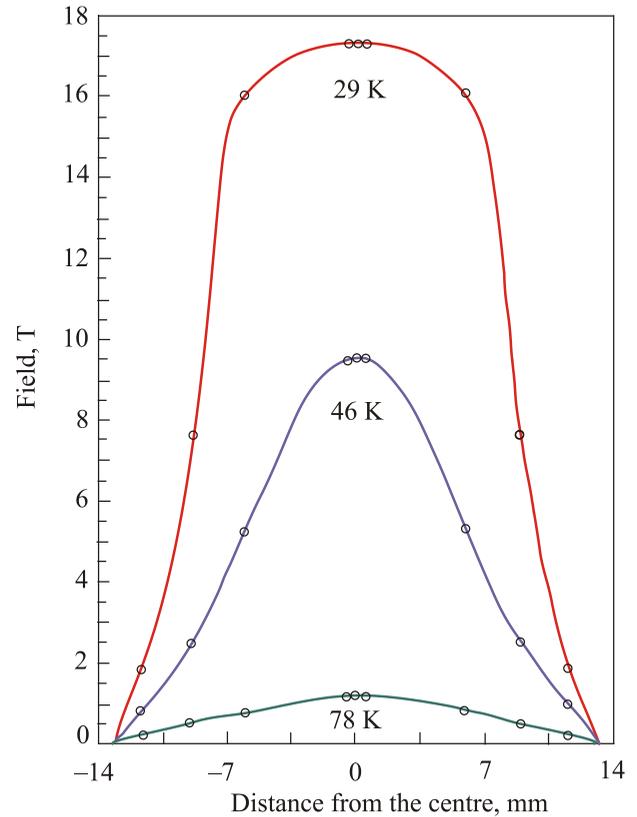

Fig. 19. The radial distributions of TMF components that are perpendicular to the surface of a YBa$_2$Cu$_3$O$_{7-\delta}$ disc (with a diameter of 28 mm), achieved at three disc temperatures (77, 46, and 29 K).[158]

In addition to the increase in the TMF in the small gap between two discs,[126,158–164] in comparison to the TMF at the surface of one disc, it was found that when several discs are stacked on top of one another the TMF component perpendicular to the disc surface also increases due to the decrease in the demagnetizing factor that occurs with the increase in sample "thickness." Thus, the TMF value in the FC mode at the surface of one GdBaCuO disc with a diameter of 140 mm and thickness of 20 mm at $T = 65$ K was $B_{tr} = 3.66$ T,[124,125] in the small (3 mm) gap between two discs $B_{tr} = 5$ T,[126] and at the surface of a system of two discs $B_{tr} = 4.3$ T.[126] It was shown in Ref. 165 that when three DyBaCuO discs with a diameter of 45 mm and a thickness of 10 mm are superimposed onto one another, the TMF at the surface of this system in FC mode at $T = 77$ K increases by 1.7 times, however once a fourth disc is added the TMF is lower than in the case of three discs. In Ref. 166 the authors put 6 discs with a diameter of 45 mm and thickness of 15 mm together. At a temperature of 38.1 K they obtained a TMF on the surface of the system $B_{tr} = 5.07$ T when applying an external field of 6 T in the FC mode. In addition to being studied in an array of discs, TMF is also studied in "brickwork" type structures composed of several layers of GdBaCuO plates $80 \times 80 \times 30$ mm in size.[167] The increase in TMF is important to the practical use of bulk HTSC samples with TMF as "permanent magnets." Examples include magnetic separators, magnetic cushion transport, i.e., any case demanding a high magnetic induction, and the current HTSC sample technology allows them to be produced mainly in the form of discs with relatively small thickness.

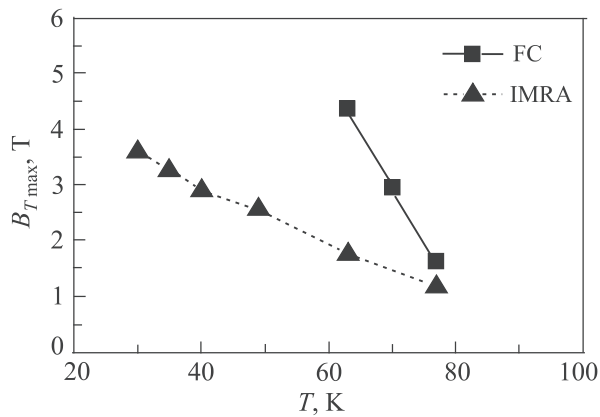

Fig. 18. The temperature dependences of the maximum trapped magnetic field for the same SmBaCuO sample magnetized using the IMRA and FC methods.[154]

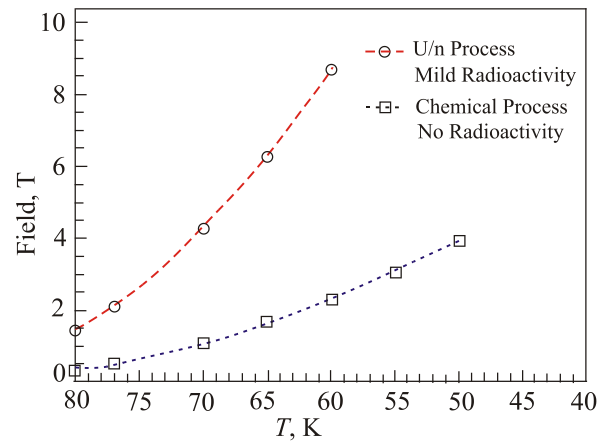

Fig. 20. The TMF temperature dependences in the FC mode for irradiated and unirradiated YBa$_2$Cu$_3$O$_{7-\delta}$ samples.[162]

It is necessary to apply double the external magnetic field to the sample in the ZFC mode as opposed to the FC mode, in order to obtain the same magnitude of the captured magnetic field.[151,152]

The amplitude of the trapped magnetic field (TMF) $B_{tr}$ both in the FC and the ZFC modes depends on the volume of the bulk HTSC and its critical current density $j_c$:[151,152,158]

$$B_{tr} = A\mu_0 j_c d, \tag{20}$$

where $A$ is the geometric factor, and $d$ is the diameter of the grain (sample).

Taking into account Eq. (20) it is necessary to cool the sample to lower temperatures, however as the temperature drops the HTSC heat capacity decreases and pinning losses rise.[168] When using refrigeration machines the low thermal conductivity of HTSCs must also be taken into account,[152,158] because it leads to jumps in the magnetic flux. When designing and manufacturing HTSC samples it is necessary to consider the strength characteristic of the HTSC. In the process of growing melted samples microcracks and cavities are inevitable. If the HTSC sample is placed in a high magnetic field, then it will be impacted by tensile stress (due to the interaction between the external magnetic field and the superconducting current) or magnetic pressure $\sigma \sim B_{tr}^2$.[158,159,164,169] which will lead to an increase in the cracks and ultimately result in sample breakage. For YBCO $\sigma_{max} = 25$–30 MPa.[154,169] Several methods are used in order to circumvent the rupture: adding silver to the process of preparation for bulk samples,[159–161] using a metal tie band (rings of stainless steel,[124–126,153,154,159–161,170,171] as in Fig. 21, or of an aluminum alloy[163,166]) steeping with adhesive in vacuum,[156,158,172] and hardening using carbon fibers or fiberglass.[156,158]

### 5.3.1. Local TMFs in the $YBa_2Cu_3O_{7-\delta}$ compound

The second, and in many ways new, direction in studies pertaining to the magnetic properties of HTSCs by means of trapping a field and developing methods of its application, is the formation of a local TMF (LTMF)[173–176] using a local external field in flat $YBa_2Cu_3O_{7-\delta}$ samples. This direction is of interest for noncontact determination of the local magnetic properties of truly inhomogeneous HTSC samples, as well as for creating magnetic memory elements based on the formation and dynamics of separate microscopic regions with a trapped field. One way to form a LTMF is by applying the action of a magnetic field generated by two microsolenoids (MS) to a HTSC plate or film located in the gap between them. The small (with a diameter of about 0.5 mm) regions with LTMFs that are created by this method at different points along the HTSC surface contain useful information as to the dependence of the LTMF ($H_{f,l}$) on the value of the local excitation ($H_{e,l}$), the cooling mode, as well as about the homogeneity of the local critical current in the sample.

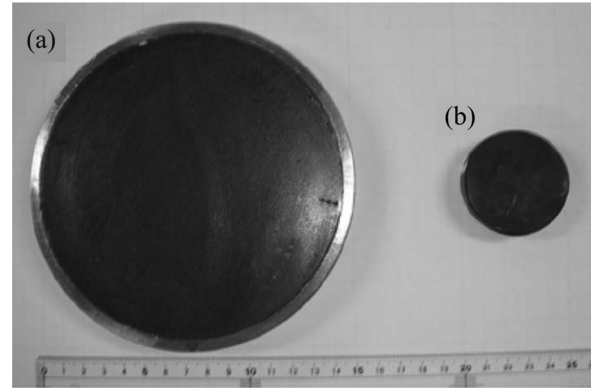

Fig. 21. The surface view of bulk HTSC discs: a GdBaCuO disc with a diameter of 140 mm is reinforced with a steel ring (1), and a disc with a diameter of 46 mm without a ring.[124]

The initial linear growth of the LTMF in the FC mode (Fig. 22), starting from the zero value of the magnetic field, points to the complete absence of the Meissner effect in the Josephson medium of the investigated granular ceramics and the full Meissner effect in the ceramics' granules at the indicated values of the external magnetic field ($H_{e,l} > H_{c1g,l} \approx 2000$ Oe). The parallelity of the linear sections of the dependences obtained in both cooling regimes suggests that after the critical field at which trapping starts $H_{c1,l} \approx 80$ Oe is attained in the ZFC mode, the process of trapping the field in both modes is the same. Thus, in the case of an FC mode the circuits formed by granules and the weak bonds between them, up to LTMF saturation ($H_{e,l} < 80$ Oe), mostly have currents that are less than the critical values, and the mechanism responsible for trapping the excitation field after it is turned off is the magnetic flux conservation law in a superconducting circuit. The ZFC mode is subject to the critical state model, according to which practically all circuits in the region at excitation fields exceeding $H_{c1J,l}$ are in a critical current state at the moment of freezing (before the excitation field is turned off), and that after the excitation field is turned off the circuits trap the difference between the critical and supercritical magnetic flux. At the same time the fact that the saturation of the LTMF at 150–160 mOe is the same for both trapping modes attests to the fact that the maximum possible critical current is achieved in all Josephson junctions formed by granules and the weak bonds between them, which limits further growth of the LTMF. The magnitude of the scattering field at the detector location depends on the size of the magnetic detector, its distance from the source of the trapped field, and LTMF microstructure in the trapping region. Our experimental values can be affected by the first two factors, and much less by the third, all of which decrease the value of the scattering field at the location of the ferroprobe. Indeed, it can be seen that in the ZFC mode with a local excitation field of $H_{e,l} = 160$ Oe, after it is turned off a field at a level of $H_{f,l} = 140$–150 mOe is captured, i.e., the scattering field of a region with LTMF is 1000 times smaller than the one that was fed. The true value of the LTMF in the sample is much higher than the value of the scattering field measured by the ferroprobe (0.4 mm in diameter and about 4 mm in length) over the LTMF region, and is comparable to the external local excitation field $H_{e,l}$, up to the saturation region of the dependence $H_{f,l}(H_{e,l})$ in the FC mode. The value of the measured scattering field over the LTMF region is therefore proportional to the LTMF inside the sample.

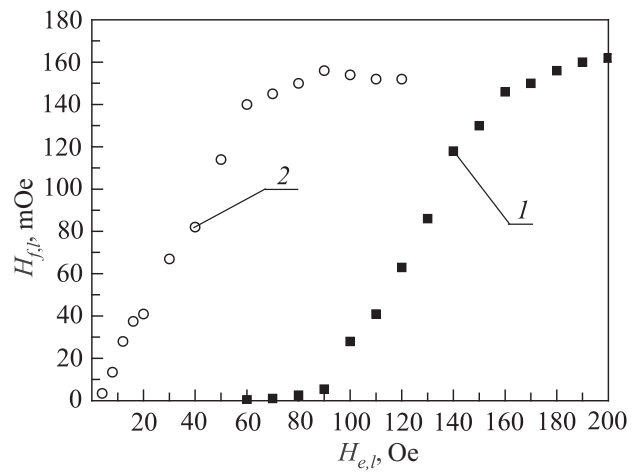

Fig. 22. Typical dependences $H_{f,l}$ ($H_{e,l}$) for the trapping region 0.5 mm in ZFC (curve 1) and FC (curve 2) cooling modes, $T = 77$ K.[174]

The dependences shown on Fig. 22 indicate that the local critical field at which the Josephson vortices start to penetrate the ceramic ($H_{c1J,l} \approx 80$ Oe) in the ZFC mode is much larger than in the case of using an external magnetic field that is uniform over the sample region (Fig. 17, where it is visible that $H_{c1J,l} \approx 2.5$ Oe). Similarly, a significantly higher value of the external local field $H_{t,l}$ (about 160 Oe) is observed at the LTMF saturation point in comparison to a homogeneous external field ($H_{t,l} \approx 30$ Oe). The differences are explained by the differences in the demagnetization coefficients of the trapped field regions when the capture is performed using a homogeneous and local field.

Figure 23 shows the dependences of the LTMF on the external local excitation field obtained at three different points along the granular ceramic plate YBa$_2$Cu$_3$O$_{7-\delta}$ in the ZFC mode at $T = 77$ K. In particular it is clearly visible that lower LTMF saturation values correspond to smaller values of the critical field and therefore to the local critical current. The differences in the dependences indicate that the superconducting properties of ceramics are not inhomogeneous. Thus, it is shown that the spatial distribution of the LTMF can serve as the basis of a new noncontact method of controlling[175] the quality of HTSC ceramics.

An area with a weak LTMF ($H_f$) can be moved along the ceramic sample using the Lorentz force $F_L$ as a constant current $I_{tr}$ is passed over the sample

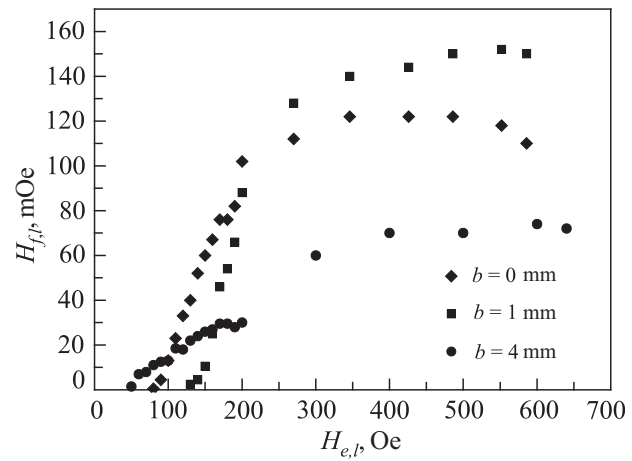

Fig. 23. The dependences $H_{f,l}(H_{e,l})$ for a LTMF with a diameter of 0.5 mm in three regions of YBa$_2$Cu$_3$O$_{7-\delta}$ granular ceramic plates that are inhomogeneous with respect to their superconducting properties, with distance from the center equal to $b = 0$, 1 and 4 mm, $T = 77$ K.[175]

$$\mathbf{F}_L \propto [\mathbf{I}_{tr} \times \mu_0 \mathbf{H}_f]. \qquad (21)$$

The collective motion of Abrikosov vortices in a type-II superconductor in a mixed state happens in a similar fashion, when under the action of the transport current. Figure 24 shows two layouts of the experiment from Ref. 176 with respect to the displacement of the LTMF region along the $X$ axis. The Lorentz force $F_L$ is generated by the interaction of the transport current $I_{tr}$ and the frozen magnetic field $H_f$.

Figure 25 shows the distributions of the vertical component of the scattering field over the LTMF region before and after passing a transport current of 3 A in different directions and a current of 5 A across the ceramic plate. It can be seen that a current of 3 A leads to the displacement of the curve maximum by about 1 mm. Changing the direction of the current leads to a displacement of the maximum by 1 mm in the opposite direction. This corresponds to the formula for the Lorentz force vector (21). In addition to the displacement of the LTMF under the action of the Lorentz force, it was found that after passing a 5 A transport current through the sample a large part (about 80%) of the weak frozen magnetic flux is removed from the sample, and that the smaller portion remains in its initial condition. Thus, the dependence after the passage of a 5 A current corresponds to the residual and practically non-displaceable part of the TMF. First, this may indicate that a part of the magnetic field is frozen on ceramic defects (pores, normal regions, etc.) through which current does not flow, and that the Lorentz force is equal to zero. Second, the observed non-displaceable part of the LTMF after the passage of a 5 A current can be related to the freezing of the magnetic field in the ceramic granules, in which case higher currents are required in order to remove the LTMF, such as $j_{cg} \gg j_{cJ}$. Recording the scattering field along the entire length of the sample instead of only in the freezing region with a diameter of about 0.5 mm is justified by the spatial extent of the TMF source scattering field, and the length of the ferroprobe detector (4 mm) in the direction perpendicular to the plane of the sample. In particular, it was found from experimental data at $T = 77$ K that the local pinning force is about $10^{-7}$ N, and the local viscosity of LTMF motion at an average velocity of $10^{-3}$ m/s was about $10^{-4}$ N s/m for a ceramic plate with an area of $10 \times 10$ mm, thickness of 0.5 mm, and a freezing zone diameter of 0.5 mm.

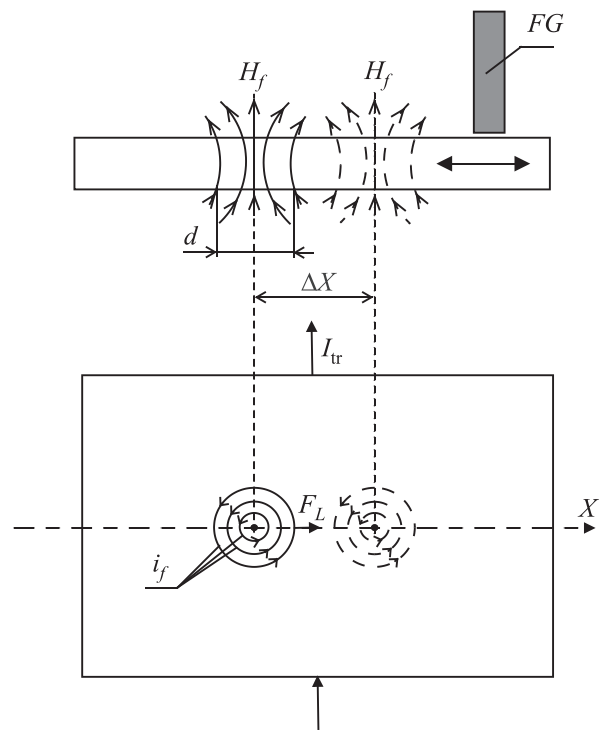

Fig. 24. A layout of an experiment involving a ceramic plate with a locally frozen field $H_f$ in a region with a diameter $d$, that can be displaced by $\Delta X$ via the Lorentz force $F_L$ ($i_f$ is the frozen vortex current, FG is the probe coil used to measure the LTMF).

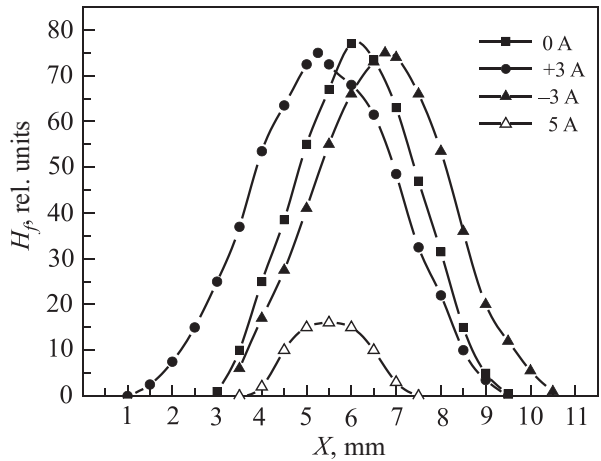

Fig. 25. The distribution curves of a frozen magnetic field near the surface of the sample along the $X$ axis of the ceramic plate (see Fig. 24) at $T = 77$ K. The inset shows the corresponding values of the current through the plate.[176]

The process of annihilation (mutual destruction) of two vortex structures in the form of LTMFs with opposite field directions,[176] which gradually approach each other under the action of oppositely directed Lorentz forces, was studied for the first time using the LTMF displacement technique. The experiment layout and the distribution of the vertical components of the scattering field over both regions with LTMFs during their transport current driven convergence are shown on the left-hand side of Fig. 26. It can be seen that when the LTMF regions merge there is a gradual mutual weakening of their LTMFs until complete disappearance, i.e., total annihilation. As the magnitude of each LTMF decreases the Lorentz force at constant current also decreases. Therefore, in order for their convergence to continue, the current must be increased. The experiment made it possible to determine exactly how much the transport current has to be increased in order for the displacement of the regions with a decreasing value of each of the LTMFs to be the same (0.5 mm). This can be seen on the right-hand side of the figure as the dependence of the required current on the distance between the maxima of the LTMF distribution. The dependence of the maxima on the distance $\Delta X$ between them is also given. Additional studies are required in order to explain its non-linear shape.

To conclude this subsection, it should be noted that it was also established that for weak fields ($H_{c,J} < H_{c1g}$) the relaxation rate of the LTMF is much less than that of a homogeneous TMF having the same magnitude.[177,178] A similar picture was observed for LTMF values greater than $H_{c1g}$.[179] As such, the high stability of the LTMF that is removed from the boundaries of the sample supports the hypothesis for the instability of a homogeneous frozen field, stating that it relaxes mainly due to a decrease in its border portion. This speaks to the fact that using the LTMF is very promising in the creation of permanent HTSC magnets.

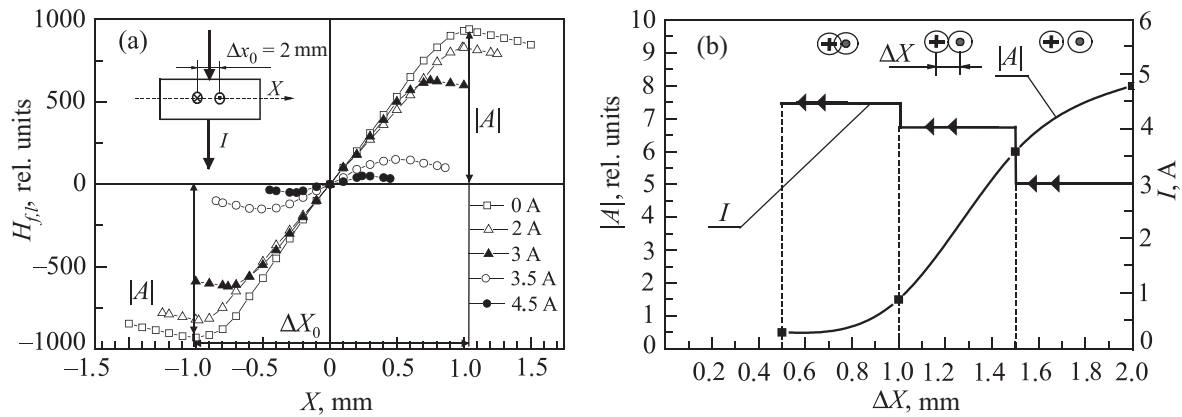

Fig. 26. Changes to the distribution of the vertical component of the scattering field at the surface of the ceramic plate (shown on the inset), along the $X$ axis of two regions with opposite LTMFs after a brief transmission of current $I$ with a magnitude of 2; 3; 3.5; 4.5 A. $\Delta x_0 = 2$ mm is the initial distance between the extrema of distribution (a). The dependence of the current $I$ that is required to displace the regions by the same distance (0.5 mm) as they converge, and the dependence of the maximum absolute values $|A|$ of both LTMFs in relative units on the distance between them, $T = 77$ K (Ref. 176) (b).

### 5.4. HTSC Josephson junctions and quantum interference properties

Earlier we considered studies that pertained mainly to issues of high-current high-temperature superconductivity. Now let us consider research in the field of low-current HTSCs. Josephson junctions and devices based on them, such as HTSC SQUIDs, are the leading representatives of this field. The Introduction mentions that the first and still to be perfected HTSC Josephson junctions (JJ) and SQUIDs made from granular ceramics of the $YBa_2Cu_3O_{7-\delta}$ compound[3–8,178] were created and in use as far back as 1987. Further efforts of the researchers were directed at creating JJ films and SQUIDs with regular JJs. The main requirements of JJs composed from HTSC materials are as follows: (a) hysteresis-free current-voltage (I-V) curves, (b) realization of the value $V_m = I_c \times R_N$ ($I_c$ is the critical junction current at $T = 77\,K$, $R_N$ is the normal junction resistance at $T = 77\,K$) at a voltage of $V_0$ that corresponds to the HTSC bandgap at $T = 77\,K$), (c) a low level of critical current fluctuation and junction resistance, (d) long-term (years) critical current stability, (e) workability of the JJ manufacturing process. Three of the main HTSC features complicate the ability to fulfill these requirements: a coherence length that is anisotropic and significantly shorter than that of LTSCs ($\sim$2 nm in the $ab$ plane and $\sim$0.2 nm in the direction of the $c$ axis); the need to ensure epitaxial film growth; the high sensitivity of the electronic properties of the barrier separating the superconductors in the HTSC junction to atomic-scale chemical and structural inhomogeneities.

Multi-year experimental studies have shown[181] that four types of HTSC JJs with predetermined parameters (Fig. 27) are the most promising.

The first type of junction (usually from a $YBa_2Cu_3O_{7-\delta}$ film) is a planar structure presenting itself as a bicrystal grain boundary (GB) between two single crystal films, the $ab$ planes of which are rotated inside the plane by a certain angle $\theta$ [as shown in Fig. 27(a)] with respect to each other. In international scientific literature this type of junction is abbreviated as BGBJ (Bicrystal Grain-Boundary Junction).

Note: if in order to ensure high-current superconductivity and a high critical current density it is necessary that the angle $\theta$ be minimal (less than 10), then in the case of a BGBJ it should be the opposite—the angle must be much larger (usually $24°$, $30°$ or $36°$) so as to ensure a low critical current density that corresponds to Josephson junctions ($10^2$–$10^3$ A/cm$^2$). A feature of this type of HTSC junction is the possibility that it could have a spontaneous jump in the phase difference of the Cooper pair wave functions, that is equal to $\pi$, in the absence of current. This possibility arises given a certain angular relationship between the positions of the $ab$ planes of the junction edges. These junctions are called $\pi$-junctions. The physical cause behind the existence of these junctions is a feature of the order parameter (OP), i.e., the bandgap of $YBa_2Cu_3O_{7-\delta}$ superconductors. If the Cooper pairs in the LTSC have an $s$-symmetry of the OP, then in these new superconductors they have a $d$-symmetry of the OP.[182] This is why this type of superconductivity is also referred to as "unusual." For example, if we consider Cooper pair wave packets propagating in mutually perpendicular directions of the HTSC crystal's $ab$ plane, then it turns out that they cancel each other[183] (Fig. 28).

On a microscopic level of the electron pairing process the theory of $d$-symmetry of the OP differs from the $s$-symmetry theory in that it allows for the existence of nonzero orbital angular momentum of Cooper pairs.[183]

One of the experiments in favor of OP $d$-symmetry in $YBa_2Cu_3O_{7-\delta}$ is the observation of an unusual magneto-field dependence of the critical current in so-called angular JJs (Fig. 29) in Ref. 184.

As can be seen in Fig. 29 the angular junction is formed simultaneously on two sides of a HTSC single crystal, the directions of which coincide with the directions of the single crystal's $ab$ plane. If the one-sided HTSC JJ has a traditional Josephson junction Fraunhofer dependence with a critical current maximum in a zero magnetic field, then for the angular junction in a nonzero field the critical current will have a minimum. These experiments are, in the opinion of a number

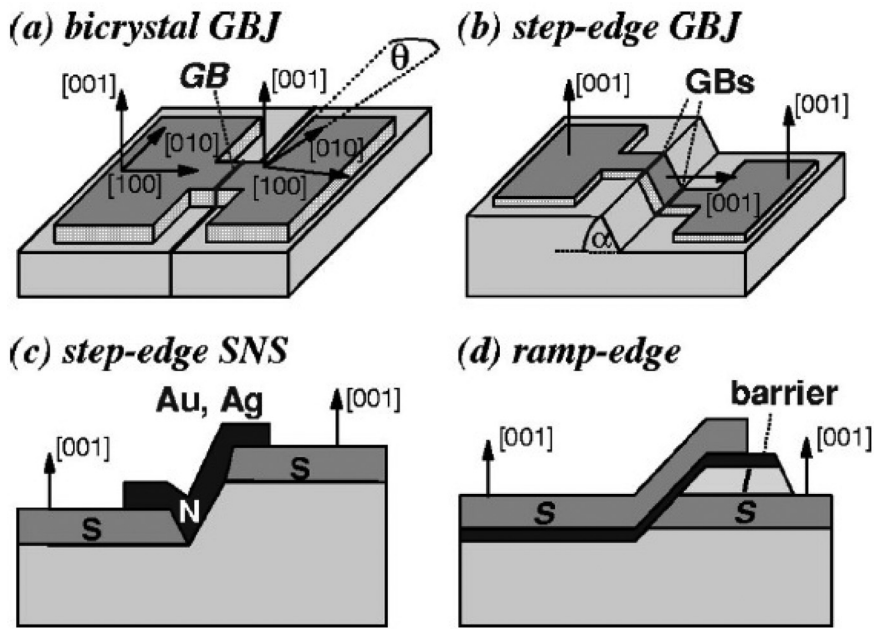

Fig. 27. Layouts of the main types of HTSC Josephson junctions: planar bicrystal junction (BGBJ) (a), bulk step-edge junction (b), bulk step-edge S-N-S junction (c), bulk ramp-edge S-I-S junction[181] (d).

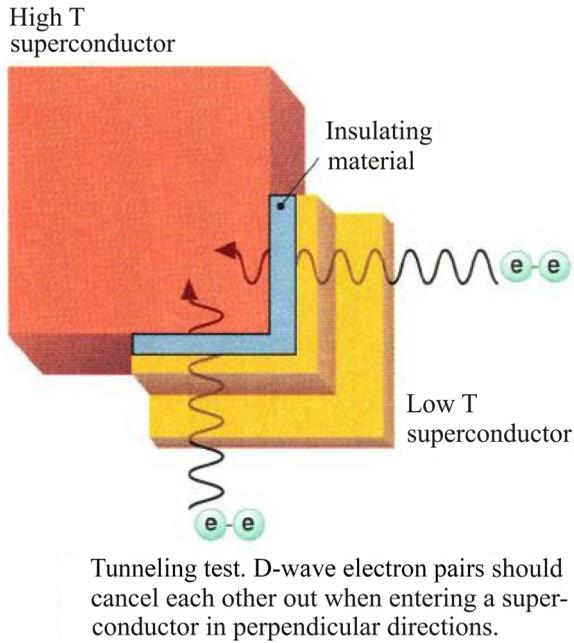

High T superconductor

Insulating material

Low T superconductor

e–e

e–e

Tunneling test. D-wave electron pairs should cancel each other out when entering a superconductor in perpendicular directions.

Fig. 28. A diagram that explains the unusual nature of the interaction between two Cooper pair wave packets and their mutual damping during their propagation along mutually perpendicular directions due to the angular Josephson junction with the HTSC crystal.[183]

of researchers,[185,186] the most substantial evidence of the idea that $d$-pairing exists in YBa$_2$Cu$_3$O$_{7-\delta}$.

Another experiment in favor of $d$-symmetry of the OP in HTSCs is the unusual quantization of magnetic flux in an interferometer with a $\pi$-junction. If in an interferometer with a traditional LTSC JJ the quantization period is equal to the flux quantum $\Phi_0$, then in this case it is equal to half the flux quantum $\Phi_0/2$. This effect is, in the opinion of the authors of Ref. 187, one of the confirmations of the theory that $d$-pairing exists in this type of superconductor. A similar magnetic flux quantization effect was observed in Ref. 188 on a much more complex planar structure of a quantum interferometer with three HTSC JJs (Fig. 30).

Typical values of $V_m$ for BGBJ are 0.1–0.4 mV. This is much smaller than $V_0$, the calculated value of which for YBCO is about 20 mV. These junctions have hysteresis-free I-V curves, and satisfactorily retain their properties under temperature cycling and prolonged (several months) storage at room temperature if they are in a special hermetic housing. Since their main field of application are SQUIDs that are based on them, BGBJ noise characteristics will be discussed below in the section about HTSC SQUIDs. The notion of manufacturability includes not only the complexity of the technological process involved in creating the junction, but also the cost of the utilized materials. In this respect BGBJ are expensive, as determined mainly by the high cost of single-crystal substrate plates (such as strontium titanate, for example) used for spraying the YBa$_2$Cu$_3$O$_{7-\delta}$ films.

The rest of the three types of junctions can be conditionally referred to as a family of step-edge junctions (SEJ). Diagrams of these types of junctions are shown in Figs. 27(b)–27(d). Figure 27(d) shows that the disorientation of the $ab$ planes in the single crystal YBa$_2$Cu$_3$O$_{7-\delta}$ films occurs on two lines of inflection of the film in the step region. The lower inflection has the lowest critical current. It provides for the Josephson properties of the junction as a whole. For this type of junction $V_m = 0.1$–0.2 mV, and it is easier to manufacture than BGBJ since it requires a single-crystal substrate with only one orientation. More detailed information about this type of HTSC junction can be obtained in Ref. 20.

The third type of HTSC junction is shown on Fig. 27(c) and is called a SEJ (S-N-S) junction, i.e., a junction between the $ab$ planes of two regions of HTSC film that is formed on the step of the substrate using a normal metal film (Au or Ag). The high resistance superconducting junction having the required short length is formed at the point of contact between the HTSC film and the normal metal. The physics of this junction have been insufficiently studied, and a significant scatter of the $I_c$ value in different samples of these S-N-S junctions is observed. The values of $V_m$ can reach 1 mV.[181] This hinders the widespread application of this type of HTSC junction. Finally, Fig. 27(d) shows the diagram of a fourth type of junction. This type is referred to as the ramp-edge Josephson junction. It is most similar to the classic LTSC Josephson junction such as S-I-S where I is the potential barrier in the form of a thin insulator layer. In a HTSC junction of this sort the insulator is usually a thin $PrBa_2Cu_3O_7$ film that is 20 nm thick.[189] The superconducting properties of these junctions largely correspond to the classic RSJ (Resistively Shunted Junction)[190] model of a Josephson junction. The value of $V_m$ in these junctions is 0.2 mV. It is considered that according to their set of parameters these junctions are most suitable for HTSC SQUIDs.[181]

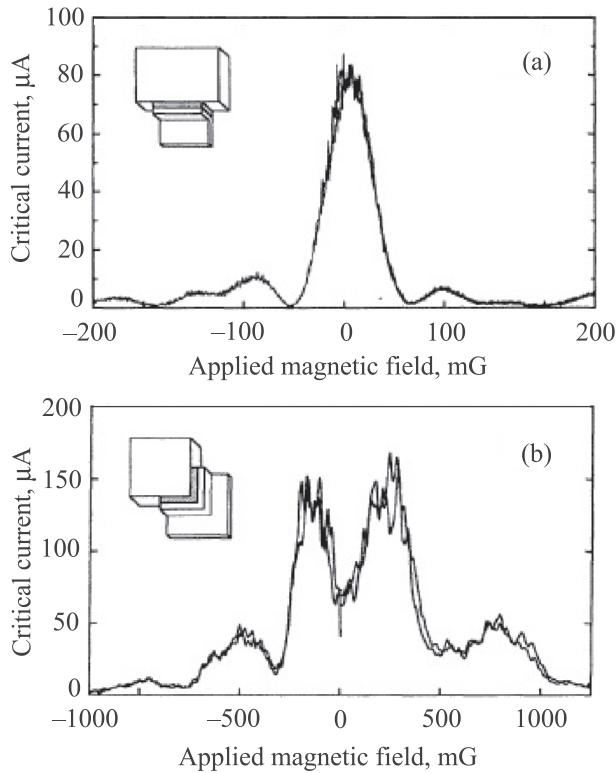

Fig. 29. Magneto-field dependences of a traditional (top) and angular (bottom) Josephson junction with an HTSC crystal at $T = 77$ K.[184] The failure of the critical current at zero field corresponds to the mutual damping of the Cooper pair flows in accordance to Fig. 28.

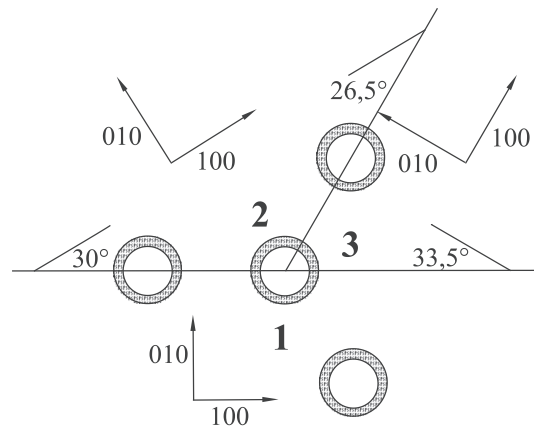

Fig. 30. A diagram of four superconducting HTSC rings, in three of which there are two or three Josephson junctions formed at bicrystal boundaries. The directions of the crystal axes of crystals 1,2,3 are shown. The junction at the boundary of crystals 2 and 3 is a $\pi$-junction. The quantization of the flux with a period of $\Phi_0/2$ (Ref. 188) is observed only on an interferometer with three JJs.

Based on the above it can be seen that values of $V_m$ that are significantly lower than the gap voltage $V_0$ are characteristic for all types of HTSC junctions prepared based on $(RE)Ba_2Cu_3O_{7-\delta}$ films. According to existing concepts[181] this is due to the imperfection of modern junction manufacturing techniques. This is expressed through the presence of localized high-density states in the insulating barrier and in the transitional S-N regions of the junction, which form an internal shunt that decreases the value of $V_m$. In addition the capture and subsequent transition of the charge carriers to another state leads to the fluctuation in the height of the local barriers, which is expressed via increased noise of the junctions and those SQUIDs that are based on them.

Summarizing the properties of HTSC junctions, it can be concluded that values of $V_m$ ranging from 0.1 to 1 mV are achieved. These values are two to 20 times higher than the known values of $V_m$ for the most common types of LTSC junctions at $T = 4.2$ K. For practical applications, mainly for the production of film SQUIDs, it is important to have a high ratio of $V_m$ to the value of intrinsic electromagnetic noise $v_n$ of the SQUID. If we assume that the gain in $V_m$ is one order of magnitude for a HTSC SQUID, then the loss in white noise (at frequencies of 10 or more hertz) is at least fivefold, Fig. 31.[181]

Therefore we can assume that when measuring the useful signals in the specified frequency range using the HTSC SQUID, the device is not inferior to modern commercial LTSC SQUIDs. In the infra-low frequency range (less than 1 Hz) LTSC SQUIDs have an undoubted advantage in sensitivity. Therefore, considering the cheaper cryogenic equipment for HTSC SQUIDs, their use is advisable for purposes of magnetocardiography, electromagnetic prospecting of minerals, non-destructive electromagnetic methods of material testing, and in scanning magnetic microscopes where useful signals at frequencies in the region of 10 Hz and above can be recorded.

To conclude the description of HTSC Josephson junctions and HTSC SQUIDs it should be noted that attempts to detect the internal Josephson effect (IJE) in single crystals of the $YBa_2Cu_3O_{7-\delta}$ family have yet to be successful, although this effect is perfectly observable in bismuth-based cuprate samples.[191]

In conclusion of Sec. 5 we should also note that the measured values and temperature dependences of the first and second critical magnetic fields of these HTSCs are quite well (as far as practical application) described by existing superconductivity theories (GL theory, as well as BCS theory for superconductors with strong interaction). In particular, calculations using the GL theory give the following parameters for $YBa_2Cu_3O_{7-\delta}$, for optimal (OD) and weak (WD) doping: the coherence length in the $ab$ plane $\xi_{ab}(0) = 1.3–2.8$ nm, in the direction of the axis with $\xi_c(0) = 0.15–0.7$ nm and magnetic field penetration depth $\lambda_{ab}(0) \approx 140$ nm, $\lambda_c(0) \approx 700$ nm,[133] which is close to the corresponding values obtained from the experiments in Refs. 134 and 135. At the same time, there are a number of important properties that cannot be explained using the known theories and demand further theoretical research. At present the microscopic theory of superconductivity for compounds considered in the review is under development. The progress is hindered by the complexity of their chemical and structural composition, the unusual superconducting properties of the compounds that are very different from the properties of earlier discovered superconductors, as well as by the strong impact that even small quantities of foreign element impurities have on the superconducting properties. We will remark on only some of these unsolved problems. These compounds, as is the case for other cuprate HTSCs, have a so-called unusual superconductivity. Its uniqueness, in comparison to those metal LTSCs that were identified before 1986 and were explained by the BCS theory within the framework of electron-phonon interaction, manifests itself in a number of important experimentally observed features. First of all, this is expressed through the weak oxygen isotope effect.[192] The BCS theory has so far failed to explain the high critical temperature of HTSC and its large changes in response to small variations in oxygen doping, considering this isotope effect. Another unusual property is the appearance of a pseudogap (PG) in cuprate HTSCs at $T^* \gg T_c$.[193,194] Below the PG temperature $T^*$ there is a significant decrease in the electron density of states at the Fermi level[195,196] for reasons that are yet to be understood. As a result, below $T^*$ there is a change to all HTSC properties[197] and the Fermi surface is most likely transformed.[198,199] Another special feature is the unusual magnetic flux quantization with a period equal to $0.5\Phi_0$, as exemplified in cuprate rings containing three Josephson junctions, formed at the joints of HTSC crystals with different orientations in the $ab$ plane (see the text above). The ratio between the bandgap and the critical temperature of these HTSCs[6,200] becomes another unusual feature when compared to BCS theory. In particular, for $YBa_2Cu_3O_{7-\delta}$ $2\Delta/kT_c$ = 4.5–5 instead of the BCS theory value of 3.5. At the same time, intensive theoretical studies based on a large amount of experimental data obtained during the 30 years that HTSCs have existed have made it possible to make significant progress in understanding the processes that occur in high-temperature superconductors. For example, the singlet pairing of charge carriers in the considered compounds,[201] the hole nature of the conductivity, the substantial contribution of the strong electron-phonon interaction[202] at least in some of the high-temperature superconductors,[203] the need to account for magnetic fluctuations[204] associated with the antiferromagnetic nature of the base compound $YBa_2Cu_3O_6$ when constructing a theory describing unusual superconductivity, as well as the existence of $d_{x2-y2}$ symmetry of the order parameter in these superconductors, can all be considered as being established.

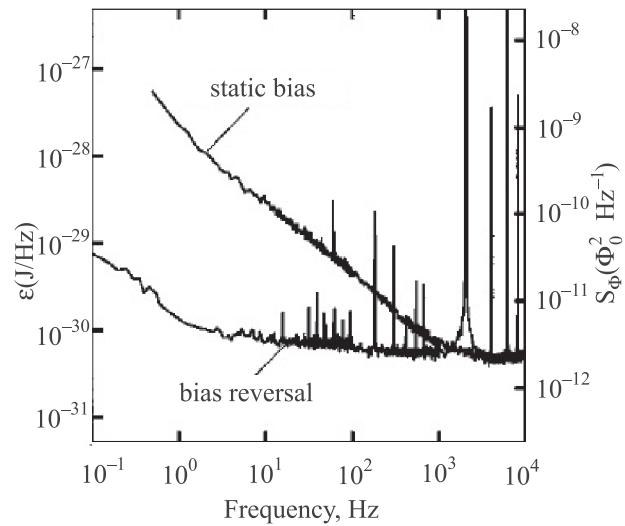

Fig. 31. Typical energy ($\varepsilon$) and magnetic (in units of spectral density of the magnetic flux noise $S_\Phi$) dependences of HTSC junction noise on the frequency. The top curve corresponds to the traditional method of supplying DC current to the SQUID, and the bottom one corresponds to the most advanced modern impulse method of current supply.[181]

Studies on cuprate HTSCs continue. In particular, extensive information about the properties and existing ideas as to the nature of the unusual superconductivity in HTSCs is given in a recent review authored by Paul Chu.[205]

## 6. Application of (RE)Ba$_2$Cu$_3$O$_{7-\delta}$

### 6.1. SQUID magnetometers

The first practical applications were found in HTSC SQUID magnetometers. On the one hand, this was made possible by the fact that cryogenic support for SQUID function became cheaper, with a transition from the liquid helium boiling point to the boiling point of liquid nitrogen, and in a some cases, to miniature and relatively cheap gas machines. On the other hand, this is also the successful development of HTSC SQUIDs based on $YBa_2Cu_3O_{7-\delta}$ films with a sensitivity that is also only slightly inferior to helium HTSC SQUIDs. Some of their drawbacks include the need to complicate their design in order to protect the HTSC film of the SQUID detector from moisture in the surrounding atmosphere and the devices' low temporal stability after storage at ambient temperature. The scope of their use includes both traditional directions involving LTSC SQUIDs, as well as new fields. Well-known manufacturers of HTSC SQUIDs and magnetometers that are based on them include American Superconductor in the USA, STAR Cryoelectronics in Germany, and Institut fur Festkorperforschung, Forschungszentrum Julich GmbH, in Julich.

### 6.2. HF filters

Narrow-band HF filters in radio technology have also been known since low-temperature superconductors first came into use, because of their high quality factor that makes it possible to separate multiple communication channels at different frequencies, while excluding their mutual influence. The reasons for the development and propagation of HTSC filters are the same as those in the case of SQUIDs. Studies pursuant to their application and manufacture based on $YBa_2Cu_3O_{7-\delta}$ films are being conducted in the USA, Russia, Germany, Japan, and China.[206]

### 6.3. Current limiters

Superconducting short-circuit current limiters for power transmission lines in high-current devices appeared as a result of HTSCs being discovered. They have advantages over traditional mechanical switches because they increase the safety of the extracurrent interruption process in the circuit and sharply reduce the overvoltage at failure. Most of the existing HTSC limiters are extended (several meters) high-temperature superconductor sections made of thick $YBa_2Cu_3O_{7-\delta}$, located in a special cryostat with liquid nitrogen. This section is included in the failure of the protected line. As line current increases above the nominal value the limiter goes into a resistive state and provides a large additional load resistance for the station generator. The line current decreases within a few milliseconds. This is enough for the low-power mechanical switch to break the line without causing overvoltage. Existing HTSC limiters are available in the USA, Germany, and Japan.[207]

### 6.4. Cables (HTSC wires)

Superconducting cables that are hundreds of meters long and can support currents in the hundreds of amperes appeared as an economically advantageous element of power transmission lines only after HTSCs were discovered. Usually flat multi-layer ribbons consisting of a steel tape, several intermediate coatings, and a $YBa_2Cu_3O_{7-\delta}$ film were used as conductors.[56] HTSC cables are manufactured in by Bruker and European Advanced Superconductors in Germany, by American Superconducting Corporation (AMSC), SuperPower Inc., in the USA, by Fujikura and Sumitomo Electric Industries Ltd., in Japan, by Nexans Superconductors in France, by Superconductor, Nano & Advanced Materials Corporation (SuNAM Co.) Ltd., in South Korea, and by Beijing Eastforce Superconducting Technology Co. in China.

### 6.5. Magnets

Figure 32 shows the external appearance of an HTSC magnet for a magnetic separator. Recently the windings of such magnets have usually been made of several flat coaxial sections of HTSC ribbon based on $YBa_2Cu_3O_{7-\delta}$ film. The magnetic field in the center of the magnet is usually several tesla at a magnet operating temperature of 25–30 K. The magnet does not use liquid cryoagents and is cooled by a gas refrigeration machine.[208]

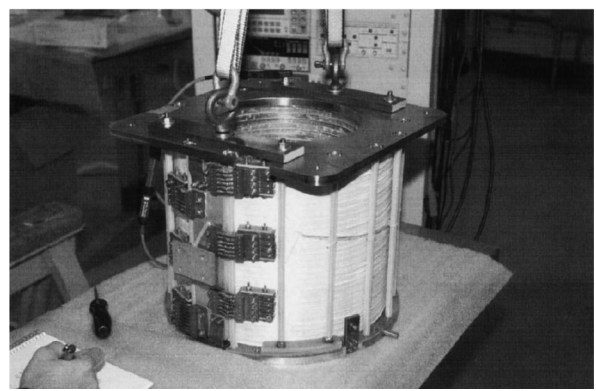

Fig. 32. The external appearance of one of the HTSC magnets used for a magnetic separator.[208]

Areas of application for such magnets, excluding multipurpose separators (enrichment of various ores and coal, purification of sewage and industrial waters, purification of raw materials in pharmaceutical production, etc.) include perspective HTSC NMR tomographs as well as magnets used for scientific research. Magnets for electric cars (engines and generators) are also promising.[24,208–212] The leading countries in which HTSC magnets are being developed are the USA, Germany, Japan, and recently China.

### 6.6. Magnetic levitation

The appearance of large-sized quasi-single crystals of compounds from the REBCO family, which was discussed in Sec. 4, stimulated the active development of the field of magnetic levitation in Japan, China, Brazil.[24,209] In addition to the "display-only" applications of this effect, one of which is shown in Fig. 33,[121] China in particular implemented a state program (863_CD080000, 1997) to create prototypes of magnetic levitation vehicles based on high-temperature superconductors. A transportation device that was capable of moving cargo of up to 800 kg for a distance of up to 100 m at a speed of 40 km/h was successfully created and tested. The main superconducting elements of the device included 100 cylindrical $YBa_2Cu_3O_{7-\delta}$ single crystals with a diameter of 30 mm and a height of 15 mm, cooled by liquid nitrogen. The lifting force of each cylinder was 90 N. The current model of one of the Japanese levitating transport devices is shown in Fig. 34.[210]

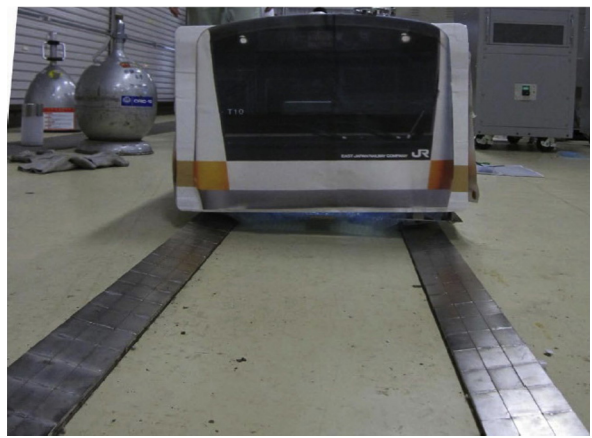

Fig. 34. The current model of a Japanese magnetic levitation transport vehicle based on a high-temperature superconductor.[210]

A promising target of the programs in China and Brazil is the creation of an ultra-high-speed transport ground vehicle moving in a vacuum tube at a speed of 1000 km/h.

A number of countries are also developing magnetic bearings for various mechanisms based on HTSC magnets.[211]

## 7. Conclusion

High-temperature superconductors are currently the main objects of research in the field of fundamental and applied superconductivity. The most famous among these HTSCs is the compound $YBa_2Cu_3O_{7-\delta}$, with a critical temperature of about 90 K. In recent years superconducting parameters of this compound and other compounds similar to it, that involve the replacement of yttrium by rare-earth elements such as Nd, Gd, Sm, Dy, Ho, have been significantly improved. Unlike early modifications to granular $YBa_2Cu_3O_{7-\delta}$ ceramics with unreproducible parameters, in recent years highly stable samples of these compounds have been developed, including bulk single-domain discs and long wires with high superconducting parameters. This was achieved by improving the technology involved in their manufacturing process via the methods of texturing and melting ceramics. The appearance of the $(RE)Ba_2Cu_3O_{7-\delta}$ HTSC sharply expanded the field of practical application for superconductivity (superconducting cables, magnetic levitation, magnetic separators, line current limiters, SQUID magnetometers that function using cheap liquid nitrogen, as well as others).

Research, theoretical models, and developments pertaining to the $YBa_2Cu_3O_{7-\delta}$ compound have served as a stimulus for work in the field of other families of cuprate HTSCs having higher critical temperatures based on thallium and mercury,[24,129] and have also facilitated research in the field of new promising iron-containing HTSCs. The achieved results have brought us substantially closer to the era of superconductivity at room temperature.

The authors wish to thank A. N. Omelyanchuk, Yu. A. Kolesnichenko, A. L. Solovjov for the useful discussions involving the content of the review, as well as V. A. Shklovsky and V. A. Finkel for kindly providing the publications related to the subject of the review.

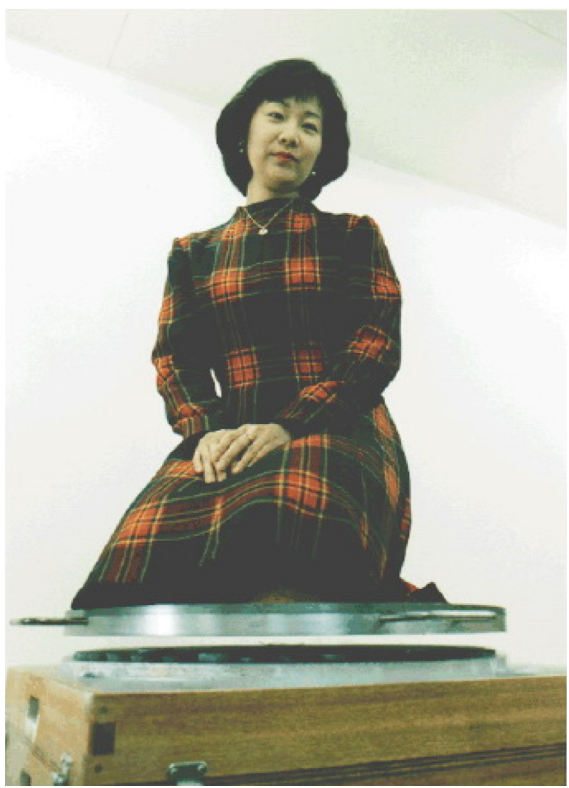

Fig. 33. A Japanese demonstration of the magnetic levitation of a disc having a built-in permanent magnet, over HTSC elements cooled by liquid nitrogen.[121]

a)Email: bondarenko@ilt.kharkov.ua